\documentclass[12pt]{article}
\usepackage[top=1in, bottom=1in, left=1.in, right=1.in]{geometry}
\usepackage[english]{babel}
\usepackage{atbegshi,cite}
\usepackage{amsmath,amssymb,amsbsy,amstext, amsthm, simplewick}
\usepackage{hyperref}
\usepackage{graphicx}
\usepackage{amsfonts}
\usepackage[small]{caption}
\usepackage{upgreek}
\usepackage[titletoc]{appendix}
\usepackage{setspace}
\usepackage[usenames,dvipsnames,table]{xcolor}
\usepackage{slashed}
\usepackage{comment}
\usepackage[normalem]{ulem}

\usepackage{caption}
\usepackage{subcaption}

\newcommand{\newc}{\newcommand}
\newc{\fpi}{f_{\pi}}
\newc{\etap}{\eta^{\prime}}
\newc{\llll}{\langle\lambda\lambda\rangle}
\newc{\FFd}{F^a\tilde F^a}
\newc{\qbar}{{\overline q}}
\newc{\TR}{{\rm Tr}}
\newc{\Kahler}{K\"ahler }
\newc{\Zbb}{{\mathbb Z}}
\newc{\Rt}{{\mathbb R}^3}
\newc{\Rf}{{\mathbb R}^4}
\newc{\So}{{\mathbb S}^1}
\newc{\zt}{{\mathbb Z}_2}
\newc{\RtSo}{{\mathbb R}^3\times{\mathbb S}^1}
\newc{\scriminus}{{\cal I}^-}
\newc{\scriplus}{{\cal I}^+}
\newc{\mpl}{M_p}
\newc{\Ricci}{\mathcal{R}}
\newc{\bv}{\phi}
\newc{\calU}{u}
\newc{\calK}{K}
\newc{\calUi}{u^{-1}}
\newc{\calG}{{\cal G}}
\newc{\calO}{{\cal O}}
\newc{\calQ}{{\cal Q}}
\newc{\calT}{{\cal T}}
\newc{\calI}{{\cal I}}
\newc{\calOb}{{\cal O}^\dagger}
\newc{\hphi}{{\hat\phi}}

\newcommand{\paren}[1]{\left(#1\right)}
\newcommand{\bracket}[1]{\left[#1\right]}

\usepackage{physics}
\usepackage{soul}
\usepackage{enumerate}

\theoremstyle{plain}
\theoremstyle{plain} 
\theoremstyle{plain} 
\theoremstyle{plain}
\theoremstyle{plain}
\theoremstyle{plain}

\renewcommand{\title}[1]{{\Large\bf\flushleft{#1}}\vspace*{3ex}\\}
\renewcommand{\author}[2]{{\noindent\hspace*{2.5em}\large#1}
                     \footnote{Electronic mail: $\mathtt{#2}$}\\}

\begin{document}
\begin{titlepage}
\begin{flushright}
{\large 
~\\
}
\end{flushright}

\vskip 2.2cm

\begin{center}

{\large
\bf Vacuum Decay and Euclidean Lattice Monte Carlo}

\vskip 1.4cm

{Jiayu Shen$^{a,b,c,}$\footnote{jiayus3@illinois.edu}, Patrick Draper$^{a,b,c}$, and Aida X. El-Khadra$^{a,b,c}$ }
\\
\vskip 1cm
{{\it $^{a}$Illinois Quantum Information Science and Technology Center, Urbana, Illinois 61801}}\\
{{\it $^{b}$Illinois Center for Advanced Studies of the Universe, Urbana, Illinois 61801}}\\
{{\it $^{c}$Department of Physics, University of Illinois at Urbana-Champaign, Urbana, Illinois 61801}}\\
\vspace{0.3cm}
\vskip 4pt

\vskip 1.5cm

\begin{abstract}
The decay rate of a metastable vacuum is usually calculated using a semiclassical approximation to the Euclidean path integral. The extension to a complete Euclidean lattice Monte Carlo computation, however, is hampered by analytic continuations that are ill-suited to numerical treatment, and the nonequilibrium nature of a metastable state. 
In this paper we develop a new methodology to compute vacuum decay rates from Monte Carlo simulations of Euclidean lattice theories. To test the new method, we consider simple quantum mechanical systems systems with metastable vacua. This work can be extended to Euclidean field theories, which we discuss in the Conclusions.
\end{abstract}

\end{center}

\vskip 1.0 cm

\end{titlepage}

\setcounter{footnote}{0} 
\setcounter{page}{1}
\setcounter{section}{0} \setcounter{subsection}{0}
\setcounter{subsubsection}{0}
\setcounter{figure}{0}

\section{Introduction}

The decay of a metastable vacuum state is an old and well-studied problem in quantum mechanics (QM) and quantum field theory (QFT). It is well-known how to compute the tunneling rate in QM using semiclassical  methods, and these techniques can be extended in a natural way to QFT~\cite{Coleman:1977py,PhysRevD.16.1762}. In recent years the theory of tunneling has received renewed attention~\cite{PhysRevD.95.085011,Guada:2020ihz,Ai:2020vhx,Lagnese:2021grb,Hayashi:2021kro,Matsui:2021oio,Croon:2021vtc,Ivanov:2022osf,Gould:2021ccf}.

Since the standard semiclassical analysis is performed using the Euclidean path integral, it is natural to ask whether Euclidean lattice theory can also be used to study vacuum decay. In addition to ordinary barrier penetration problems, lattice methods could be useful for quantitative studies of vacuum decay in situations where the semiclassical methods are inadequate, such as the decay of vacua that emerge from strong dynamics (see e.g. Ref.~\cite{Teper:2008yi}). Formulating and refining a lattice approach to these problems might also yield methods of more general interest and applicability.

However, Euclidean Monte Carlo (MC) simulations of false vacua are not without subtleties. A configuration which begins in a metastable state, or in a false vacuum (FV), will 
evolve in Monte Carlo time to eventually thermally fluctuate over the barrier. In the semiclassical limit, the barrier ``peak'' is a saddle point of the classical action, a solution known as the bounce~\cite{Coleman:1977py}, and the Monte Carlo time evolution can be thought of schematically as ``false vacuum $\rightarrow$ bounce $\rightarrow$ true vacuum.''
If the true vacuum (TV) is deep, as a practical matter, the system will never return to the false vacuum after thermalization, so all configurations in the thermalized ensemble describe the true vacuum.
They are exponentially more important than the bounce and they are only rendered innocuous after a final analytic continuation back to real time, a point emphasized in the study of Ref.~\cite{PhysRevD.95.085011} which sought to place the problem of vacuum decay on more rigorous footing. 
This analytic continuation is more or less straightforward in semiclassical analyses, but it is impractical in an MC approach.

In this paper, we develop a new framework to compute approximate but accurate decay rates from Euclidean lattice simulations. To test the approach, we consider QM tunneling problems as illustrated in Fig.~\ref{fig:pot1}. 
Our primary results are the definition of a new observable that approximates the decay rate of a quantum mechanical metastable vacuum, a prescription for its computation in Euclidean Monte Carlo simulations, and numerical simulations testing the accuracy of the method.

The remainder of this paper is organized as follows. In Sec.~\ref{sec:theory} we develop the necessary theoretical tools, define our computational approach, and describe the systematic uncertainties introduced by the associated  
approximations. In Sec.~\ref{sec:examples} we  apply the method to a representative family of potentials. An advantage of studying QM tunneling problems is the ability to compute the decay rate by solving the time-dependent Schr\"odinger equation (TDSE). 
We perform three-way comparisons between results obtained from solving the TDSE (``exact''), from Euclidean lattice Monte Carlo computations (``lattice''), and from semiclassical analyses. We find good agreement between the results over several decades in the decay rate, thus establishing the accuracy of our lattice method. 
In Sec.~\ref{sec:longlifetime} we turn our attention to very long lifetimes, where computing the rate from ensembles of practical sizes requires a different approach. We propose the ``constrained ensemble reweighting'' method and illustrate it with an example. Our conclusions are presented in Sec.~\ref{sec:concl}, where we further outline how our framework can be extended to Euclidean quantum field theories.

\section{Vacuum Decay in Euclidean Lattice Theory\label{sec:theory}}

\subsection{Preliminaries\label{sec:preliminaries}}

\begin{figure}[ht!]
\begin{center}
\includegraphics[width=0.6\linewidth]{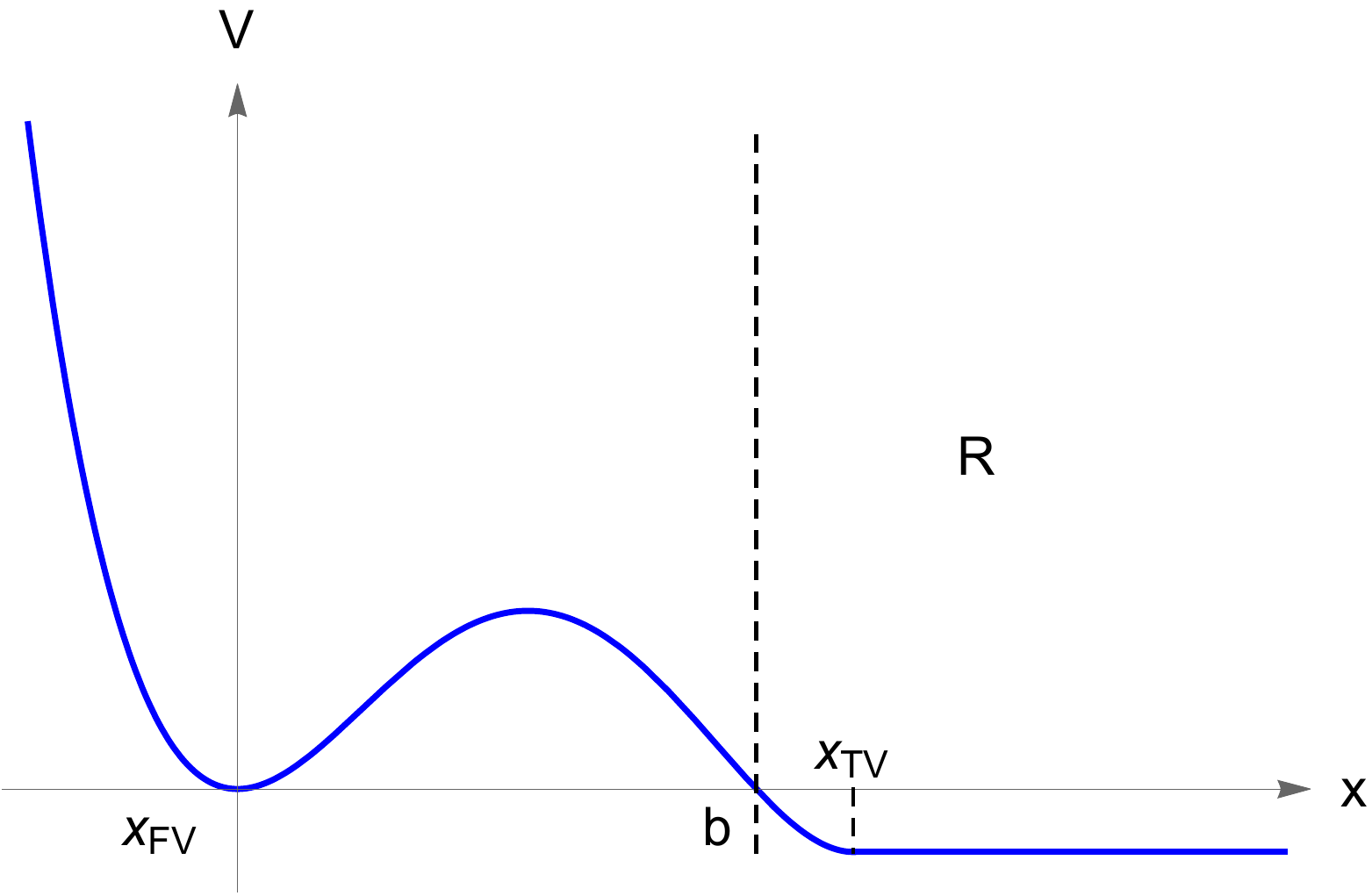}
\caption{Example potential $V(x)$. $x_{\mathrm{FV}}$ is  the local potential minimum corresponding to the false vacuum. $x_{\mathrm{TV}}$ is  the starting position of a global-minimum plateau region of the potential. $b$ is the classical turning point that satisfies $V(b) = V(x_{\mathrm{FV}})$. $R = \{ x \; | \; V(x) < V_{\mathrm{FV}} \} = \{ x \; | \; x > b \}$ is the classically allowed region.} 
\label{fig:pot1}
\end{center}
\end{figure}

We consider single-particle quantum mechanics with a tunneling potential. An example potential is shown in Fig.~\ref{fig:pot1}.  
The continuum Euclidean action  is
\begin{equation}
    S_E = \int \, dt \left(\frac{1}{2} \paren{\frac{d x}{d t}}^2 + V\paren{x} \right).
\end{equation}
In this normalization $x$ is treated as a $0+1$D field: the kinetic term has a dimensionless coefficient $1/2$, so that the dimension of $x$ is $[x] = [E^{-1/2}]$. This definition of $x$ is used throughout this  paper. 
With the false vacuum positioned at $x_\mathrm{FV} = 0$, we parametrize the leading term in the expansion of the potential around $x_\mathrm{FV}$ as $V(x) = \frac{1}{2} m^2 x^2 + \ldots$. Since this term has the same form as the mass term in scalar field theories, we can consider the dimensionful parameter $m$ as the mass of the particle. 
A more detailed description of the potential is given in Sec.~\ref{sec:params}.

The continuum Euclidean path integral facilitates a convenient semiclassical treatment of  false vacuum decay. One first constructs the bounce, a solution $x_b(t)$ to the Euclidean equations of motion that asymptotes to the classical false vacuum at early and late times. The leading order (LO)  decay rate is governed by the bounce action, $\Gamma \sim e^{-S_E[x_b]}$. The next-to-leading-order (NLO) correction is given by  the quadratic fluctuation integrals around the bounce. In these integrals the low lying modes of the fluctuation operator must be treated separately. Zero modes associated with symmetries can be treated with a collective coordinate method. More importantly, the bounce is always associated with a single mode of negative eigenvalue. The integral over the amplitude of this mode is divergent and is generally defined by analytic continuation. 

On
 the lattice, a simple choice for the discretized action is
\begin{equation}
    S_{\mathrm{lat}} = a \sum_{i = 1}^{N_T} \paren{ - \frac{1}{2} x_i \frac{x_{i+1} - 2 x_i + x_{i-1}}{a^2} + V \paren{x_i}}
    ,
\end{equation}
where $a$ is the lattice spacing and $N_T = 2 T / a$ is the total number of sites ($2T$ is the total time). The difference between the lattice action and the continuum action is $O(a^2)$ due to the discrete second-order derivative.

In order to study  vacuum decay in Euclidean lattice Monte Carlo simulations, we must first identify an observable that can be related to the desired decay rate and computed with Monte Carlo methods. We show that the probability density to find the particle at the classical turning point has the desired properties and describe its computation with Euclidean path integrals and its relation to the decay rate in Sec.~\ref{sec:probabilitydensities}.

Any continuum calculation in Euclidean time must be analytically continued to real time. However, such continuations are impractical in lattice Monte Carlo computations because they require exponential sensitivity. We elaborate on the problem in Sec.~\ref{sec:cut} and define a procedure that avoids the need for analytic continuation, removing the exponential sensitivity requirement, at the cost of introducing a systematic error.

\subsection{Probability Densities from Euclidean Path Integrals}
\label{sec:probabilitydensities}

The probability density for the system to be in the state $|x\rangle$ at time $t$, given that we started from a normalized state $\psi$ at $t=0$, is
\begin{align}
\rho(x,t) = \abs{\braket{x,t}{\psi,0}}^2. 
\end{align}
When $\ket{\psi} = \ket{\mathrm{FV}}$, a metastable state localized near the classical false vacuum, the decay rate is defined as
\begin{align}
\Gamma &= - \lim_{T \rightarrow \infty} \frac{1}{P(\mathrm{FV},T)}\frac{dP(\mathrm{FV},T)}{dT},\nonumber\\
P(\mathrm{FV}, T)&\equiv \int_\mathrm{FV} dx\,\rho(x,T) = \int_\mathrm{FV} dx\,|\langle x,T|\mathrm{FV},0\rangle|^2,\nonumber\\
P(R, T) & \equiv \int_{R} dx\,\rho(x,T) = 1 - P(\mathrm{FV}, T)
.
\label{eq:defs_decay}
\end{align}
The result for $\Gamma$ should not be sensitive to the exact definition of the $\mathrm{FV}$ region, as long as it reasonably contains the point $x_{\mathrm{FV}}$ and does not extend beyond $b$.  
The long $T$ limit of Eq.~(\ref{eq:defs_decay}) is satisfied when $T$ is large compared to the ``escape attempt time'' $\sim 1/m$ in the false vacuum, $1/m \ll T$. If we consider times within the long $T$ limit that are short compared to $1/\Gamma$, then the probability $P(\mathrm{FV}, T) \approx 1$, and the decay rate can be estimated as 
\begin{align}
\Gamma\approx - {\dot P}(\mathrm{FV},T) = {\dot P} (R, T), \quad\quad 1 / m\ll T \ll 1 / \Gamma
\end{align}
in this regime.

Now let us relate $\dot P(R, T)$ to $\rho$. We have
\begin{align}
\dot P(R,T) = \int_R dx\, \dot\rho(x,T)
=j(b,T).
\end{align}
Here $j(b, T)$ is a probability current flowing through $x = b$, and we have used the continuity equation $\dot\rho(x, T)=-\partial_x j(x, T)$.
We can also define a probability flow velocity $u$ through
\begin{align}
j(x,T)\equiv u(x, T)\rho(x,T).
\end{align}
Semiclassically, the probability flow velocity can be estimated from the classical definition of the kinetic energy $E_{\mathrm{FV}} - V(x) = (1/2) u_{\mathrm{cl}}(x, T)^2$, where $E_{\mathrm{FV}}\approx (1/2)m$ is the quantum vacuum energy of the approximate quadratic potential centered at $x_{\mathrm{FV}}$. 
For $1/m \ll T \ll 1/\Gamma$ and $x = b$, we have $u(b, T) \approx \sqrt{m}$.

\begin{figure}[!t]
\centering
		\includegraphics[width=0.5\textwidth]{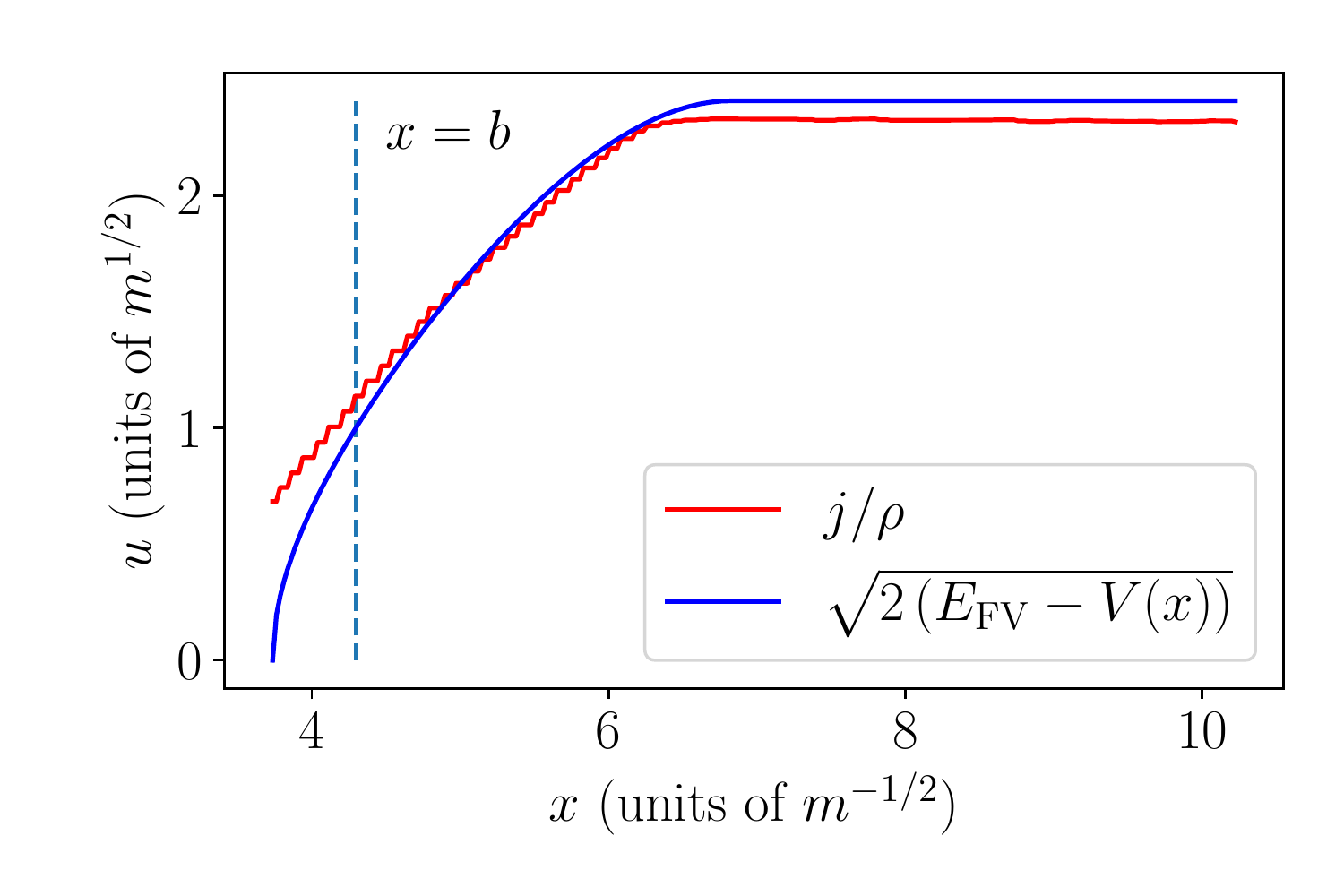}
\caption{A comparison between the probability flow velocity $u = j / \rho$ and $u_{\mathrm{cl}} = \sqrt{2 (E_{\mathrm{FV}} -  V(x))}$ for the potential of the form in Fig.~\ref{fig:pot1} [more precisely, the potential of Eq.~(\ref{eq:pw_pot2}) with $\alpha = 0.9$, $\beta = 8.0$], computed by numerically solving the TDSE. Agreement occurs for $x \gtrsim b$, and $u(b) = j(b) / \rho(b)$ is approximately equal to $u_{\mathrm{cl}}(b) = \sqrt{m}$. Only the range of $x$ with $E_{\mathrm{FV}} -  V(x) \geq 0$ near $x = b$ or in the classically allowed region is plotted.
}
\label{fig:u_comparison}
\end{figure}

The relationship $u(b, T) \approx \sqrt{m}$ is easily validated for specific examples by the numerical solution of the time-dependent Schr\"odinger equation (TDSE). In Fig.~\ref{fig:u_comparison}, we compare $u = j / \rho$ from the full quantum mechanics and the approximation $u_{\mathrm{cl}} = \sqrt{2 (E_{\mathrm{FV}} - V(x))}$, exhibiting good agreement when $x \gtrsim b$. $u_{\mathrm{cl}}$ is not expected to match $j / \rho$ in the classically forbidden region, \textit{i.e.}, when $x$ is substantially smaller than $b$.

Therefore, if $\rho(b, T)$ can be computed by other means, then the decay rate can be estimated as
\begin{align}
    \Gamma \approx {\dot P} (R, T) = j(b, T) \approx \sqrt{m} \rho(b, T).
    \label{eq:gammaappx}
\end{align}

The advantage of this formulation is that $\rho(b, T)$ can be evaluated with a Euclidean path integral and is approximately independent of $T$ in the time range of interest $1/m \ll T\ll 1/\Gamma$ described above. We define a Euclidean transition amplitude,
\begin{align}
A(\psi, b;T) =\langle b|e^{-HT}|\psi\rangle =  \int dy\,\psi(y) K (y, b; T)
,
\label{eqAdef}
\end{align}
where the Euclidean propagator over time $T$ between some $y$ and $z$ is
\begin{align}
    K (y, z; T) = \int_{x(\tau=0)=y}^{x(\tau=T)=z} \mathcal{D}x\, e^{-\int_0^{T} d\tau L_E[x]}
    .
\end{align}
The real-time probability density is 
\begin{align}
&\rho(\psi,0;b,T) \equiv |\langle b|e^{-iHT}|\psi\rangle|^2\nonumber\\
&=\int dy\,\psi(y) \left( \int_{x(t=0)=y}^{x(t=T)=b} Dx\, e^{i\int_0^T dt L[x]}\right) \int dz\,\psi^*(z)\left( \int_{x(t=0)=z}^{x(t=T)=b} Dx\, e^{-i\int_0^T dt L[x]}\right)\nonumber\\
&= \int dy\,\psi(y)  \left( \int_{x(\tau=0)=y}^{x(\tau=iT)=b} Dx\, e^{-\int_0^{iT} d\tau L_E[x]}\right) \int dz\,\psi^*(z) \left( \int_{x(\tau=0)=z}^{x(\tau=-iT)=b} Dx\, e^{-\int_0^{-iT} d\tau L_E[x]}\right)\nonumber\\
&=\int dy\,\psi(y) \left( \int_{x(\tau=0)=y}^{x(\tau=T)=b} Dx\, e^{-\int_0^{T} d\tau L_E[x]}\right)\bigg|_{T\rightarrow i T}\int dz\,\psi^*(z) \left( \int_{x(\tau=0)=z}^{x(\tau=-T)=b} Dx\, e^{-\int_0^{-T} d\tau L_E[x]}\right)\bigg|_{T\rightarrow i T}\nonumber\\
&=\left[ \int dydz\,\psi(y)\psi^*(z)K(y,b;T)K(z,b;-T) \right]\bigg|_{T\rightarrow i T}\nonumber\\
&=\left[ A(\psi,b;T)A(\psi^*,b;-T)\right]\bigg|_{T\rightarrow i T}
.
\end{align}
In the third line, we make variable changes $t = -i \tau$ in the first integral and $t= i \tau$ in the second integral. At this point there is no analytic continuation and $\tau$ is imaginary. Subsequently we analytically continue to real $\tau$ in the clockwise direction in both integrals. The ``$|_{T\rightarrow i T}$" operation denotes a counterclockwise  continuation back to Minkowski time after the integrals are computed. The Euclidean quantity $A(\psi^*,b;-T)$ is defined formally by computing $A(\psi^*,b;T)$ and continuing $T\rightarrow -T$, where $\ket{\psi^*}$ is the complex conjugate of the state $\ket{\psi}$ in the position representation. Generalizing to an unnormalized initial state $\psi$, we have
\begin{align}
\rho(\psi,0;b,T)=\left[ \frac{A(\psi,b;T)A(\psi^*,b;-T)}{\int dy\,A(\psi,y;T)A(\psi^*,y;-T)}\right]\bigg|_{T\rightarrow i T}
.
\label{eq:rhoA_1}
\end{align}
Before the replacement $T \rightarrow i T$, both the numerator and the denominator  are Euclidean path integrals and the total time extent is $2T$. 
The numerator has a path constraint $x(\tau = 0) = b$ while the denominator does not.

In the decay of a false vacuum, there is a range of Lorentzian time over which we expect $\rho$ is approximately time-independent. This occurs on timescales $1/m \ll T\ll1/\Gamma$. This is also true in the Euclidean picture if  each amplitude in the numerator of Eq.~(\ref{eq:rhoA_1}) is dominated by localized events (similar to half of a single  bounce solution, in semiclassical language), so that again changing the duration $T$ does not appreciably change the amplitude. We now make this assumption and interrogate it in Sec.~\ref{sec:numerics}. 

With both the Euclidean and Lorentzian amplitudes approximately time-independent,
the continuation in Eq.~(\ref{eq:rhoA_1}) can be ignored. 
 Any $T$-dependence in the normalization of the initial state  cancels with the $T$-dependence in the normalization of the denominator. Another way to describe this time-independence is to say that the false vacuum is  almost an energy eigenstate $\ket{\epsilon}$  of the complete Hamiltonian. 
Therefore on timescales short compared to $1/\Gamma$, the state does not change appreciably and  $\sqrt{m}\rho(b,T)\approx \Gamma$, a constant. The dominant Euclidean time evolution in $A$,
\begin{align}
    A(\psi,b;T)\approx e^{-\epsilon T}\langle b|\epsilon\rangle,
\end{align}
 cancels between the numerator and the denominator of $\rho$.

Equation~(\ref{eq:rhoA_1}) is still not in the form of an expectation value of an observable, which would be convenient for computation in Euclidean MC simulations. To relate it to such an observable, we exploit the time-independence described above and the symmetry of the Euclidean amplitudes. We consider real initial wave functions $\psi(y) \in \mathbb{R}$ and write
\begin{align}
\hat \rho(\psi,0;b,T) &\equiv \frac{A(\psi,b;T)A(b,\psi;T)}{\int dy\,A(\psi,y;T)A(y,\psi;T)}\nonumber\\
&=\frac{A(\psi,b;T)A(b,\psi;T)}{A(\psi,\psi;2T)}\nonumber\\
&=\langle f(b;0)\rangle
\label{eq:rhohatdef}
\end{align}
where
\begin{align}
    f \paren{b; 0} \equiv \lim_{\delta \rightarrow 0} \frac{1}{\delta} \Theta \paren{ \bracket{x(0) - \paren{b - \frac{\delta}{2}}} \bracket{\paren{b + \frac{\delta}{2}} - x(0)} }.
    \label{eq:fdef}
\end{align}
In practice, when $\delta$ is chosen finite and small enough, $f(b; 0)$ is an observable that returns $1/\delta$ if a path is in a small region $[b-\delta,b]$ at time $t=0$, and zero otherwise. In our calculation, we use $\delta = 0.04 \sqrt{\beta / m}$, since $\sqrt{\beta / m}$ is a characteristic scale for $x$ as is shown in Eq.~(\ref{eq:x_bar_x}).

The definition of $\hat\rho$ differs from that of $\rho$ by $T\rightarrow -T$ in the second factors of $A$ and the absence of analytic continuation of $T$. However, if
in the time regime of interest both $\rho$ and $\hat\rho$ are approximately $T$-independent, then
\begin{align}
\rho(\psi, 0; b, T) \approx \hat \rho(\psi, 0; b, T)\;\;\;\;\;(1/m \ll T\ll1/\Gamma)\;.
\label{eq:rhoequalsrhohat}
\end{align}
We examine the $T$-dependence of $\hat\rho$ below, where we find that with one important modification we can indeed approximate it as $T$-independent.

In the Monte Carlo simulation, we use periodic boundary conditions (PBCs) $x(-T) = x(T)$, $\dot{x} (-T) = \dot{x} (T)$ in Euclidean time with large $T$, so that the state $\psi$ into which the ensemble initially thermalizes is approximately the perturbative ground state in the false vacuum. This allows us to exploit time translation symmetry and improve the ensemble statistics. With PBCs, the rare events where $x(t) \gtrsim b$ can occur at a random Euclidean time $t$, and all random times have an equal chance for such rare events. Therefore, we can average the probability density at $b$ over all Euclidean times $t \in [-T, T)$ to approximate $\hat{\rho} (\mathrm{FV}, 0; b, T)$.

To summarize, we have related the decay rate to an observable $\hat\rho(b)$ that can be computed in MC. There are three primary approximations which introduce uncertainties into the result. First, we assume that $T$ can be chosen so that $1/m\ll T\ll 1/\Gamma$, which allows both the approximation $\Gamma \approx  -\dot P(\mathrm{FV})$ and the analytic continuations described above. Second, we approximate the probability flow velocity by $u\approx \sqrt{m}$, which is a fairly good approximation in practice, as we verify by explicit comparison with the TDSE solution.  Third, we assume that the relevant Euclidean amplitudes are dominated by trajectories that probe beyond the barrier in localized, rare events, so that they are insensitive to $T$. The $T$ interval $1/m\ll T\ll 1/\Gamma$ is necessary but not sufficient for this to be true, as we discuss in the next subsection.

We note that our method is complementary to the ``direct method'' of Ref.~\cite{PhysRevD.95.085011}, which is also expressed using the Euclidean path integral. The direct method involves  taking an imaginary part after the analytic continuation. Such a procedure, when applied to a Monte Carlo calculation, may be sensitive to the details of how the analytic continuation is performed. Instead, our method avoids taking an imaginary part by constructing an observable that is approximately independent of $T$, so that the analytic continuation $T \rightarrow i T$ is rendered innocuous.

\subsection{Cuts in Ensemble Generation and Postselection: Controlling the Negative Mode\label{sec:cut}}

Although we have identified a useful lattice observable, there is still an issue of the unwanted dominance of true-vacuum-like configurations in MC that must be addressed before we can apply it to real simulations. 
We now illustrate the problem in detail, using semiclassical language for convenience, and describe a practical resolution for  lattice MC computations.

Let us briefly review the NLO semiclassical contribution to the decay rate to establish notation and ideas. We decompose paths near the bounce as
\begin{align}
    x \paren{t} = x_b \paren{t} + \sum_{n=0}^{
    \infty} c_n x_n \paren{t}
    ,
\end{align}
with the normalization condition,
\begin{align}
    \int_{-T}^{T} dt \, x_m \paren{t} x_n \paren{t} = \delta_{mn}
    .
\end{align}
The basis $\{ x_n \}$ is chosen such that it diagonalizes the Euclidean action expanded to the quadratic order as
\begin{align}
    \bracket{- \frac{d^2}{dt^2} + V''\paren{x_b\paren{t}}} x_n \paren{t} = \lambda_n x_n \paren{t}
    ,
\end{align}
with the ordering of $n$ defined through $\lambda_0 \leq \lambda_1 \leq \lambda_2 \leq \cdots$.
The NLO contribution to the path integral around the bounce is 
\begin{align}
      \prod_{n=0}^{\infty} \int d c_n \exp \paren{ - \frac{1}{2} \lambda_n c_n^2}
\end{align}
up to an overall normalization.
However, the lowest eigenvalue is negative, $\lambda_0 < 0$, and the second-lowest eigenvalue is zero, $\lambda_1 = 0$. The zero-mode $x_1 \paren{t}$ reflects the time translation invariance of the bounce $x_b(t)$, so the integral of $c_1$ can be replaced by an integral of the center time of the bounce which gives a factor of $2 T$, still convergent for large but finite $T$. The negative mode $x_0(t)$  leads to an exponential divergence. Qualitatively, this divergence can be explained by the the fact that a path $x(t)$ that spends the majority of its time near the true vacuum has an action about $- 2 \abs{V_{\mathrm{TV}}} T < 0$, much lower than the bounce action $S_b = S[x_b] > 0$.

\begin{figure}[!t]
\centering
	\begin{subfigure}[ht]{0.48\textwidth}
		\centering
		\includegraphics[width=\textwidth]{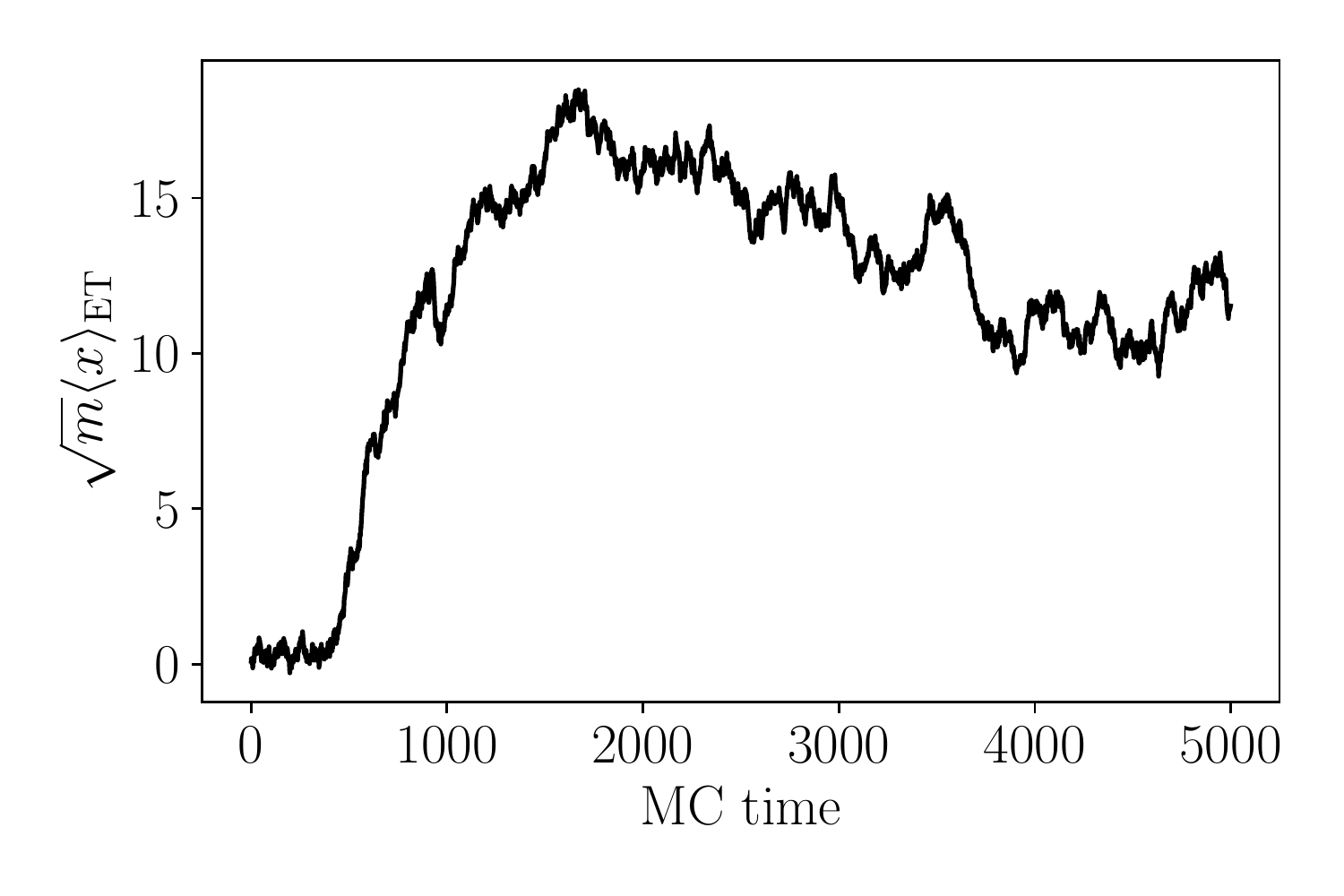}
		\caption{}
		\label{subfig:avg_x_list}
	\end{subfigure}
	\vfill
	\begin{subfigure}[!htbp]{0.48\textwidth}
		\centering
		\includegraphics[width=\textwidth]{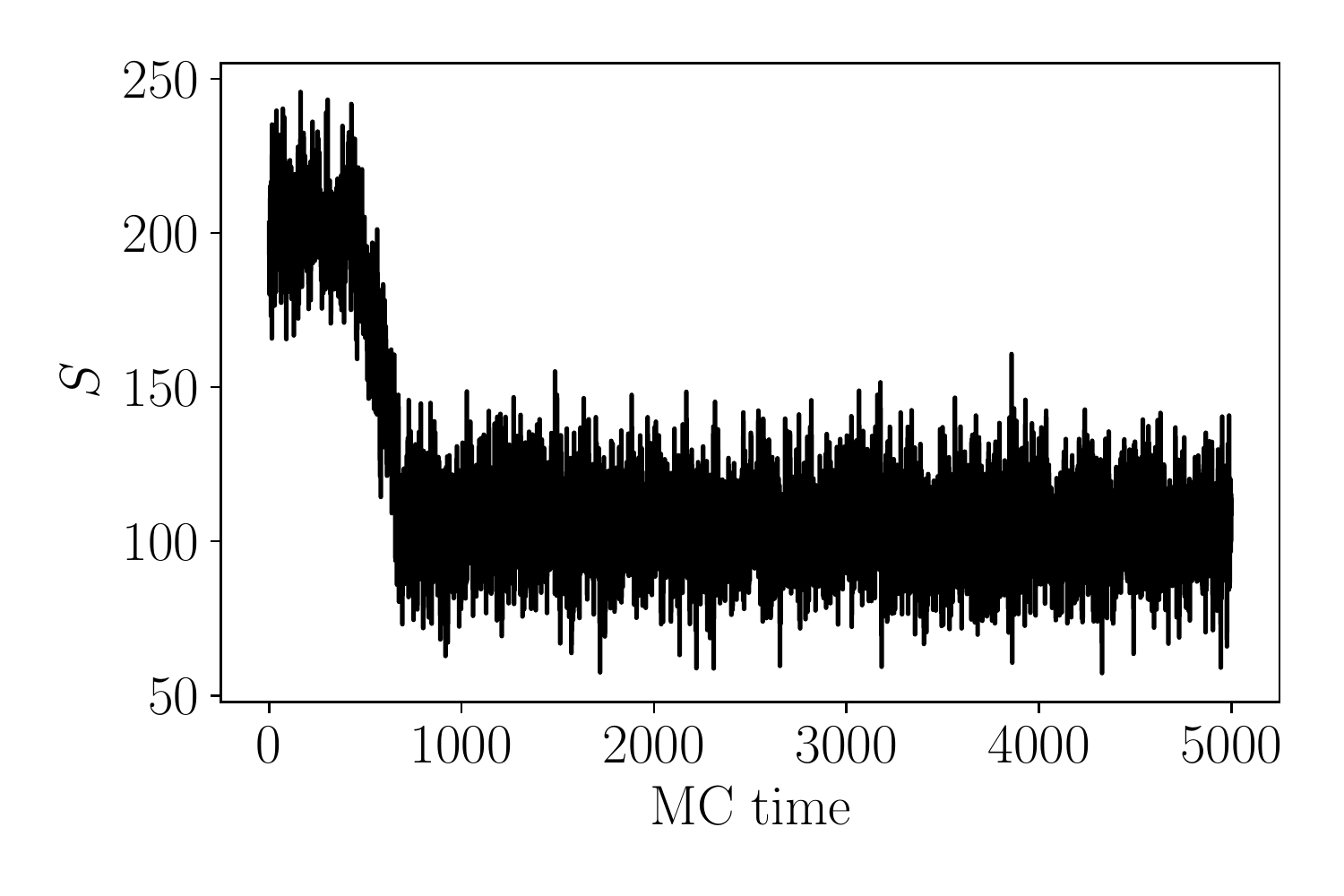}
		\caption{}
		\label{subfig:action_list}
	\end{subfigure}
\caption{Some properties of configurations as a function of Monte Carlo time in a simulation of an example potential like in Fig.~\ref{fig:pot1} (more precisely, a simulation with $\alpha = 0.9$, $\beta = 8.0$, $a = 0.1$, $N_T = 400$, in the notation of Sec.~\ref{sec:numerics}). Panel  (\subref{subfig:avg_x_list}) shows $x$ averaged over Euclidean time (ET), \textit{i.e.}, $\langle x \rangle_{\mathrm{ET}} = \int dt \, x(t) / \int dt$. Panel (\subref{subfig:action_list}) shows the Euclidean action. The initial configuration is taken to be $x(t) = x_{\mathrm{FV}}$, no thermalization steps are taken, and the adjacent MC times are relatively correlated. At early Monte Carlo times, the configuration remains near the false vacuum, with $\langle x \rangle \approx x_{\mathrm{FV}} = 0$ and a relatively high Euclidean action due to the quantum fluctuations around the false vacuum. For MC times from about 500 to 1000, a transition starts which takes the configuration from the false vacuum to the true vacuum region with $x > x_{\mathrm{TV}} = 6.82 / \sqrt{m}$ in this example. The Euclidean action after the transition is significantly lower than before the transition. In this example, the difference in action $\sim 100$ implies an enormous suppression $\sim e^{-100}$ for the ensemble to ever return to the false vacuum. In other words, if the ensemble reaches global equilibrium, then the FV-static configuration is essentially never observed, and the thermalized ensemble behaves like that of a free particle because of the flat potential at $x > x_{\mathrm{TV}}$. [The free particle has equal probability to move in both directions, which is responsible for the (random) decrease in $\langle x \rangle_{\mathrm{ET}}$ after the transition.]
}
\label{fig:mc_problematic}
\end{figure}

Typically, an analytic continuation in the $c_0$ contour is taken to make the integral converge. However, in a Monte Carlo simulation,  an analogous procedure to ``analytic continuation of the $c_0$ contour'' is not available. Once the ensemble generation passes near the bounce saddle point of the action, with high probability it will rapidly evolve toward configurations that spend most of their time near the true vacuum.
The action of a typical configuration in the above situation is then even lower than the action of a typical false-vacuum-like configuration, so it is extremely unlikely to fluctuate back over the saddle point.
This behavior is illustrated in Fig.~\ref{fig:mc_problematic}. Starting with a false-vacuum-like configuration enables the observation of two distinct perturbative vacua, but the ensemble is not useful for quantitatively computing the decay rate.

To obtain a useful result from Monte Carlo, we impose a cut to discard configurations that go too far into the direction of the true vacuum. First, let us return to the semiclassical picture and see the effect of cutting off the $c_0$ integral instead of continuing it. 

On a finite interval $c_0 \in [c_0^{\mathrm{min}}, c_0^{\mathrm{max}}] \sim (c_0^{\mathrm{min}} < 0 < c_0^{\mathrm{max}})$ the negative mode integral is
\begin{align}
\int_{c_0^{\mathrm{min}}}^{c_0^{\mathrm{max}}} dc_0\, e^{-\frac{1}{2} \lambda_0 c_0^2} &= \sqrt{\frac{\pi}{2 \left|\lambda_0\right|}} \bracket{\mathrm{Erfi}\paren{\sqrt{\frac{\left| \lambda_0 \right|}{2}} c_0^{\mathrm{max}}  } + \mathrm{Erfi}\paren{\sqrt{\frac{\left| \lambda_0 \right|}{2}} \left|c_0^{\mathrm{min}}\right|}}\nonumber\\
&\approx e^{\frac{1}{2} \left|\lambda_0\right| \left(c_0^{\mathrm{max}}\right)^2}\left(\frac{1}{\left|\lambda_0\right| c_0^{\mathrm{max}} }+\dots\right) + e^{\frac{1}{2} \left|\lambda_0\right| \left(c_0^{\mathrm{min}}\right)^2}\left(\frac{1}{\left|\lambda_0 c_0^{\mathrm{min}}\right| }+\dots\right) \pm i  \sqrt{2 \frac{\pi}{ \left|\lambda_0\right|}}
.
\label{eq:negative_mode_integral_0}
\end{align}
The constant pure imaginary term takes ``$+$'' for the integral limits deformed to $ \arg(- c_0^{\mathrm{min}}) = \arg(c_0^{\mathrm{max}}) \in (0, \pi)$ and ``$-$'' for $ \arg(-c_0^{\mathrm{min}}) = \arg(c_0^{\mathrm{max}}) \in (\pi, 2 \pi)$. At $ \arg(-c_0^{\mathrm{min}}) = \arg(c_0^{\mathrm{max}}) = 0\text{ or }\pi$, the asymptotic expansion at large $|c_0^{\mathrm{min}}|$ and $|c_0^{\mathrm{max}}|$ is ill-defined.
For a finite $T$, with a convention $x_0 (0) > 0$, the increasing direction of $c_0$ drives the configuration $x_b (t) + c_0 x_0(t)$ toward the true vacuum region $R$. In fact, in the full functional integral, when $T$ is finite, there are always effective cutoffs on fluctuations in the $c_0$ direction, and these cutoffs are proportional to $T$. For example, in the positive $c_0$ direction, the lowest possible action configuration is the true vacuum, where the action is $S_{\mathrm{TV}}=2 V_{\mathrm{TV}} T$. $c_0^{\mathrm{max}}$ is a function of $T$ with an unknown functional form, but $c_0^{\mathrm{max}} \rightarrow \infty$ when $T \rightarrow \infty$. In the other direction there is an effective cutoff associated with the false vacuum configuration. 
Therefore, if the integral in Eq.~(\ref{eq:negative_mode_integral_0}) is analytically continued by replacing $T \rightarrow i T$ and with the limit $T \rightarrow \infty$ taken before computing the integral, then the final result is $- i \sqrt{2 \pi / |\lambda_0|}$. Its imaginary part combined with the fluctuation integrals of other modes gives the NLO decay rate. This is why the continuation $T \rightarrow i T$ is both subtle and important: it removes exponentially large $T$-dependent contributions to the Euclidean amplitudes~\cite{PhysRevD.95.085011}.
However, it is impractical to numerically evaluate the path integral at  large $T$ with such high precision that the finite constant term $- i \sqrt{2 \pi / |\lambda_0|}$ can be resolved against a ``background'' term that exponentially grows with $T$. We need a more aggressive cut on configurations that fluctuate too far toward the true vacuum.

Again we begin with the semiclassical computation. When finite cuts $c_0^{\mathrm{max}}$ and $c_0^{\mathrm{min}}$ are imposed, then the $c_0$ integral is a finite number that is generically unequal to $\sqrt{2 \pi / |\lambda_0|}$, but may be close to it for a suitable choice of cuts. For example, ordinary Gaussian integrals are dominated by the region within a standard deviation or so of the peak. 
Let us therefore set
$(1/2) \lambda_0 (c_0^{\mathrm{min}})^2 + 1 = 0$. 
Then 
\begin{align}
\int_{c_0^{\mathrm{min}}}^{0} dc_0\, e^{-\frac{1}{2} \lambda_0 c_0^2} = \sqrt{\frac{\pi}{2 \left|\lambda_0\right|}}  \mathrm{Erfi}\paren{\sqrt{\frac{\left| \lambda_0 \right|}{2}} \left|c_0^{\mathrm{min}}\right|} = \sqrt{\frac{\pi}{2 \left|\lambda_0\right|}}  \mathrm{Erfi}\paren{1} = & 0.83 \sqrt{2 \frac{\pi}{ \left|\lambda_0\right|}}\nonumber \\
= & 0.83 \, \mathrm{Im} \int_{- i \infty}^{i \infty} dc_0\, e^{-\frac{1}{2} \lambda_0 c_0^2}.
\end{align}
As long as the cutoff $c_0^{\mathrm{max}}$ satisfies $(1/2) |\lambda_0|  (c_0^{\mathrm{max}})^2 \leq 1$, 
we have
\begin{align}
\int_{c_0^{\mathrm{min}}}^{c_0^{\mathrm{max}}} dc_0\, e^{-\frac{1}{2} \lambda_0 c_0^2} \in \left( 0.83, 1.67 \right) \, \mathrm{Im} \int_{- i \infty}^{0} dc_0\, e^{-\frac{1}{2} \lambda_0 c_0^2} \sim O(1) \, \mathrm{Im} \int_{- i \infty}^{i \infty} dc_0\, e^{-\frac{1}{2} \lambda_0 c_0^2}
.
\end{align}
Therefore, without continuing the contour and simply placing cutoffs on the negative mode integral, we can compute the NLO decay rate up to 
an $O(1)$ relative correction.

However, beyond the semiclassical approximation, for example in Monte Carlo simulation, it is not obvious how to implement a cut on $c_0$ when the theory is formulated in configurations $\{ x(t) \}$ instead of the $\{c_n\}$ basis. We need a different approach with similar properties. Instead, we consider a functional of $x (t)$ defined as
\begin{equation}
    S^b_V \bracket{x} = \int dt \, V \paren{ x \paren{t} } \Theta \paren{x \paren{t} - b}
    ,
    \label{eq:SbVdef}
\end{equation}
where $\Theta (\cdot)$ is the Heaviside step function. Only  times $t$ such that $x(t) > b$, i.e., the configuration goes beyond the point $b$ and into the classically allowed region $R$, contributes to $S^b_V$.  $V(x(t))$ is lower than $V(b) = V_{\mathrm{FV}} = 0$ and thus negative when $x(t) > b$. In other words, $S^b_V$ measures the contribution to the action solely from the parts that can lower it below the action of the false vacuum. Configurations can be characterized into a one-parameter family using $S^b_V [x]$. The greater $x(t)$ is, when between $b$ and $x_{\mathrm{TV}}$, the more negative $V(x(t))$ is. Therefore, the lower $S^b_V[x]$ is, the more likely the configuration $x(t)$ is to be close to the true vacuum, with a more positive value of $c_0$. Configurations near the false vacuum all have $S^b_V = 0$ since they do not enter the region $R$. Roughly speaking, $c_0$ increases when $S^b_V$ decreases.

We place a hard wall on $S^b_V [x]$ during ensemble generation, and then place a more stringent cut on it during postselection. The latter is taken at the minimum location of the probability density function $p(S^b_V)$. This corresponds to not rejecting too many configurations (cutting off the Gaussian $c_0$ integral too close to the peak) while not moving too far in the direction of the true vacuum (where the result becomes exponentially sensitive to the cutoff). At the minimum of $p(S^b_V)$, results for observables are also minimally sensitive to the precise choice of the cut. In Appendix~\ref{sec:sbv} we give a more detailed justification for this choice and test it on example potentials.

With this prescription for eliminating unwanted configurations, we anticipate that the ensembles indeed satisfy the conditions such that $\hat\rho$ is approximately $T$-independent and provides a good estimate of the rate $\Gamma$. We now turn to testing the method numerically on various example potentials.

\section{Numerical Examples\label{sec:numerics}}
\label{sec:examples}
In this section we apply the algorithm described above to a family of model potentials, comparing the results with semiclassical computations and numerical solution of the time-dependent Schr\"odinger equation.  

Because decays are generally rare events, the probability of obtaining bouncelike configurations in the Monte Carlo simulation is suppressed. In semiclassical language, the rate is exponentially small in the bounce action. If this suppression is too extreme, direct ensemble generation methods do not work. To avoid this problem, this section is focused on examples where the decay rate is not prohibitively small. The case of small decay rates is considered in Sec.~\ref{sec:longlifetime}.

\subsection{The Potentials and Semiclassical Properties\label{sec:params}}

We use ``modified double-well potentials'' of the form shown schematically in Fig.~\ref{fig:pot1} as a family of useful QM examples.   We parametrize the potential as
\begin{equation}
    V \paren{x} =  \begin{cases} 
      \frac{1}{2} m^2 x^2 - \eta x^3 + \frac{\lambda}{8} x^4 & x < x_{\mathrm{TV}} \\
      V_{\mathrm{TV}} & x \geq x_{\mathrm{TV}}
      ,
   \end{cases}
\label{eq:pw_pot1}
\end{equation}
where the value of $V_{\mathrm{TV}}$ is defined to maintain the continuity of the potential at $x = x_{\mathrm{TV}}$. (We remind the reader that in our normalization $x$ is a $0+1$D scalar field and thus has the dimension $\mathrm{energy}^{-1/2}$ rather than the dimension of a physical position, $\mathrm{energy}^{-1}$.) We define the potential so that $x_{\mathrm{FV}} = 0$ and $V_{\mathrm{FV}} \equiv V(x_{\mathrm{FV}}) = 0$.

The large flat region to the right of $x_{\mathrm{TV}}$ is useful to have a continuum or quasicontinuum of unbound states for the metastable state localized around $x_{\mathrm{FV}}$ to decay into. The classical turning point is labeled by $b$ and the classically allowed region is $R = \{ x \; | \; V(x) < V_{\mathrm{FV}} \} = \{ x \; | \; x > b \}$. In the region $x \leq x_{\mathrm{TV}}$ this potential is exactly a quartic potential, so the semiclassical analysis is very similar to the case of the latter potential.

The only three parameters in this model are $m$, $\eta$, and $\lambda$. We then reparametrize the theory using a similar parametrization as in Ref.~\cite{PhysRevD.72.125004}. 
With the nondimensionalization into $\bar{t}$ and $\bar{x}$,
\begin{equation}
    \bar{t} = m t
\end{equation}
\begin{equation}
    x = \frac{m^2}{2 \eta} \bar{x}
    ,
\end{equation}
the Euclidean action of a path $x(t)$ that does not enter the modified region $x \geq x_{\mathrm{TV}}$ can be rewritten as
\begin{equation}
    S \bracket{x} = \beta \int d \bar{t} \, \bracket{\frac{1}{2} \paren{\partial_{\bar{t}} \bar{x}}^2 + \frac{1}{2} \bar{x} - \frac{1}{2} \bar{x}^3 + \frac{\alpha}{8} \bar{x}^4}
    ,
\end{equation}
where there are two dimensionless parameters,
\begin{equation}
    \alpha = \frac{\lambda m^2}{4 \eta^2}
    ,
\end{equation}
\begin{equation}
    \beta = \frac{m^5}{4 \eta^2}
    .
\end{equation}
We then choose $m$ as the only dimensionful parameter. Thus $m$ sets the energy scales of the theory, and we mostly work in units where $m = 1$. When needed, $m$ can be restored from dimensional analysis. $m$, $\alpha$, and $\beta$ form the new set of parameters that are a rearrangement of $m$, $\eta$, and $\lambda$.

With the new parametrization, the potential in Eq.~(\ref{eq:pw_pot1}) takes the form 
\begin{equation}
    V \paren{x} =  \begin{cases} 
      \frac{1}{2} m^2 x^2 - \frac{m^{5/2}}{2 \sqrt{\beta}} x^3 + \frac{m^3 \alpha}{8 \beta} x^4 & x < x_{\mathrm{TV}} \\
      V_{\mathrm{TV}} & x \geq x_{\mathrm{TV}}
      .
   \end{cases}
\label{eq:pw_pot2}
\end{equation}
We can analytically solve for the classical vacua and turning point,
\begin{equation}
    x_{\mathrm{FV}} = 0
    ,
\end{equation}
\begin{equation}
    x_{\mathrm{TV}} = \sqrt{\frac{\beta}{m}} \frac{3 + \sqrt{9 - 8 \alpha}}{2 \alpha}
    ,
\end{equation}
\begin{equation*}
    b = \sqrt{\frac{\beta}{m}} \frac{2 \paren{1 - \sqrt{\alpha - 1}}}{\alpha}
    .
\end{equation*}
We further define the dimensionless potential,
\begin{equation}
    \bar{V} \paren{\bar{x}} =  \begin{cases} 
      \frac{1}{2} \bar{x} - \frac{1}{2} \bar{x}^3 + \frac{\alpha}{8} \bar{x}^4 & \bar{x} < \bar{x}_{\mathrm{TV}} \\
      \bar{V}_{\mathrm{TV}} & \bar{x} \geq \bar{x}_{\mathrm{TV}}
      ,
   \end{cases}
\end{equation}
the dimensionless Euclidean Lagrangian,
\begin{equation}
    \bar{L}_E \bracket{\bar{x}} = \frac{1}{2} \paren{\partial_{\bar{t}} \bar{x}}^2 + \bar{V} \paren{\bar{x}}
    ,
\end{equation}
and the corresponding action,
\begin{equation}
    \bar{S} \bracket{\bar{x}} = \int d \bar{t} \, \bar{L}_E \bracket{\bar{x}}
    .
\end{equation}
This action is dependent only on $\alpha$ and independent of $m$ and $\beta$.
The complete action is proportional to $\beta$,
\begin{equation}
    S \bracket{x} = \beta \bar{S} \bracket{\bar{x}}
    .
\end{equation}
Some useful relations between the two parametrizations are
\begin{equation}
    x = \sqrt{\frac{\beta}{m}} \bar{x}
    ,
\label{eq:x_bar_x}
\end{equation}
\begin{equation}
    V \paren{x} = m \beta \bar{V} \paren{\bar{x}}
    ,
\end{equation}
and the $n$-th-order derivative of $\bar{V}(\bar{x})$,
\begin{equation}
    V^{\paren{n}} \paren{x} = m^{\frac{n}{2} + 1} \beta^{1 - \frac{n}{2}} \bar{V}^{\paren{n}} \paren{\bar{x}}
    .
\end{equation}
Therefore, $V_{\mathrm{TV}} = m \beta \bar{V}_{\mathrm{TV}}$.

The parameter $\alpha$ always satisfies $0 < \alpha < 1$ and controls the shape of the potential. In the limit  $\alpha \rightarrow 1$, the false and true vacua become degenerate as $V_{\mathrm{TV}} \rightarrow - 2 m \beta (1 - \alpha)$. In the limit  $\alpha \rightarrow 0$, the true vacuum approaches  minus infinity with  leading behavior $V_{\mathrm{TV}} \rightarrow - 27 m \beta / (8 \alpha^3)$. $\beta$ is always positive and controls the overall scale of $V$. $\beta \rightarrow \infty$ is the semiclassical limit where the quantum theory is governed by the classical bounce solution $x_b (t)$ (saddle point). Effects from quantum fluctuations $\delta x (t) = x (t) - x_b (t)$, except for the negative and zero modes, are exponentially suppressed by $\exp(-\beta (\bar{S} [\bar{x}_b + \delta \bar{x}] - \bar{S} [\bar{x}_b]))$ when $\beta$ is large.

\subsection{Simulation Results}

After introducing the $S_V^b$ cut described in Sec.~\ref{sec:cut}, we can perform a lattice Monte Carlo computation of $\hat{\rho} (\mathrm{FV}, 0; b, T)$, \textit{i.e.}, the probability density at $x=b$ at Euclidean time $T$ starting from the false vacuum state at time zero. We impose  periodic boundary conditions in Euclidean time to improve the statistics; for large $T$, the temperature is low enough that the system initially thermalizes close to the false vacuum if the Markov chain is seeded with an initial configuration equal to the semiclassical false vacuum, $x=0$.

To establish an appropriate cut on $S_V^b$, we first compute the probability density function $p(S_V^b)$. To find the minimum of this function, a finite sample may not be sufficient, since the function exhibits statistical fluctuations and we are interested in the region where $p(S_V^b)$ is approximately flat. 
We use  kernel density estimation (KDE)~\cite{10.1214/aoms/1177704472,10.1214/aoms/1177728190} and gradient descent to compute $p(S_V^b)$ and search for the minimum. We use the Epanechnikov kernel~\cite{doi:10.1137/1114019} with the kernel width small enough to capture local variation of the density function but still large enough to contain sufficient configurations. The typical scale of the kernel width for our setup is $O(10^{-1})$. In each iteration step, KDE can compute $p$ at the target $S_V^b$ from the gradient descent with a low cost. In the gradient descent method, we start from several initial values of $S_V^b$ and compare the local minima found by different initial values, due to statistical fluctuation, to find the global minimum.

As shown above, the decay rate $\Gamma \approx \rho_b (T)$ when $1/m \ll T \ll 1/\Gamma$. Therefore, we report $\rho_b$ computed from MC as $\Gamma$ and compare it against $\Gamma$ computed from the solution of the TDSE, the NLO semiclassical Gel'fand-Yaglom (GY) method, and the LO semiclassical/dimensional analysis (DA) method $\Gamma_{DA} = m e^{-S_b}$. The TDSE and semiclassical results are only expected to agree in the far  semiclassical limit, and comparing both with the MC results provides a measure of how much information the MC can access beyond the different levels of semiclassical approximation in intermediate regimes.

\begin{figure}[!htb]
\centering
	\begin{subfigure}[ht]{0.48\textwidth}
		\centering
		\includegraphics[width=\textwidth]{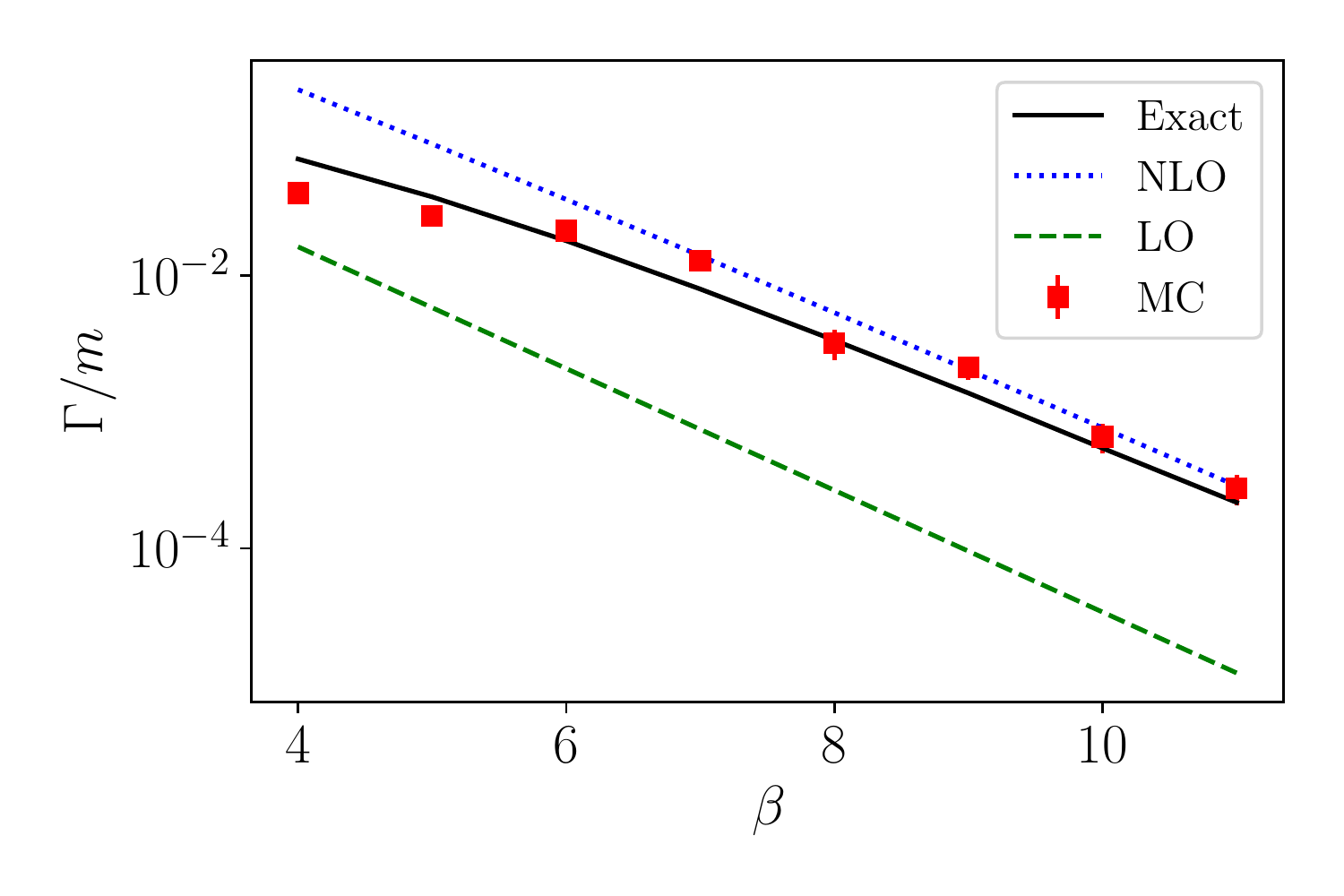}
		\caption{}
		\label{subfig:Gamma_vs_beta_alpha_0.900}
	\end{subfigure}
	\hfill
	\begin{subfigure}[!htbp]{0.48\textwidth}
		\centering
		\includegraphics[width=\textwidth]{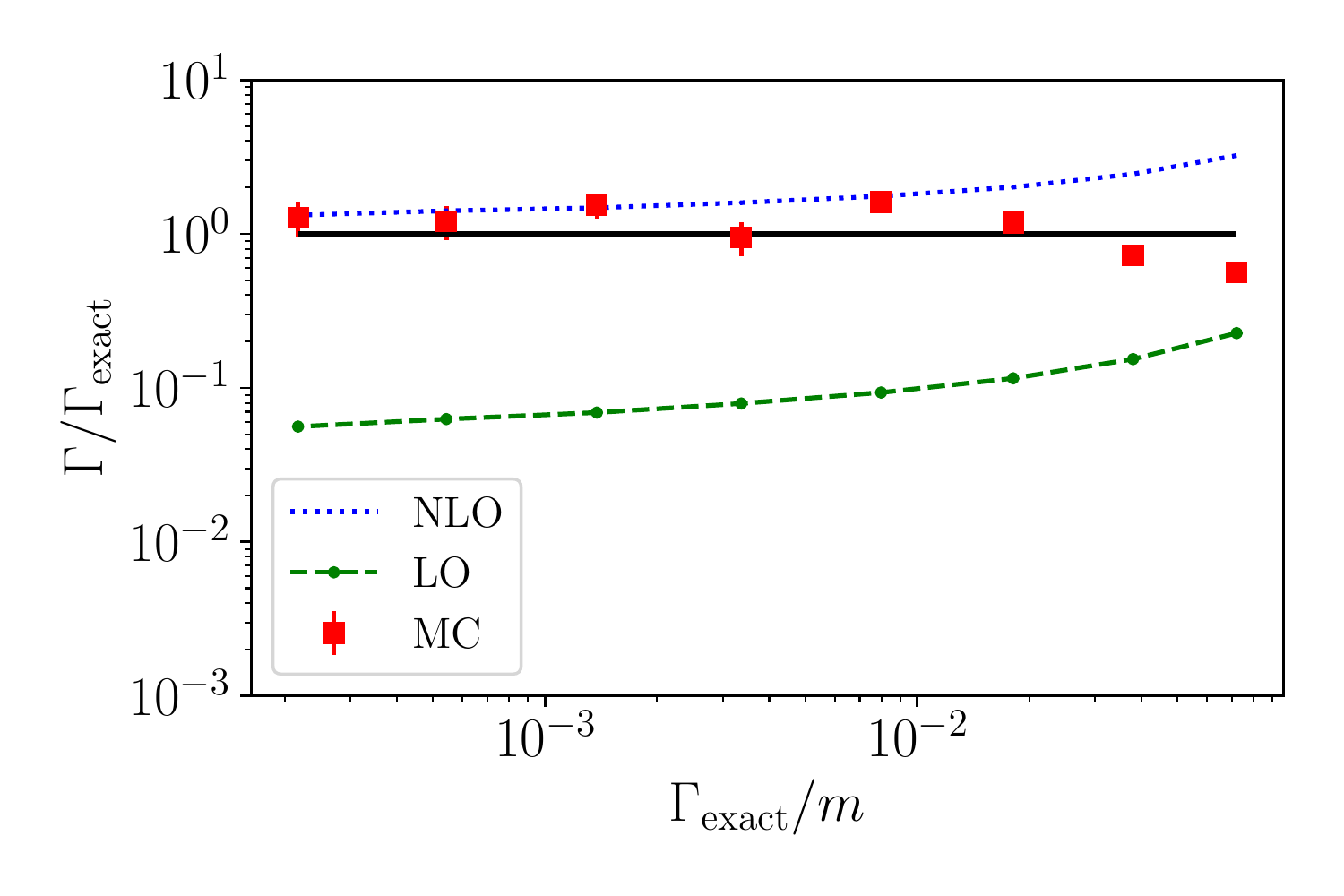}
		\caption{}
		\label{subfig:Gamma_ratio_vs_GammaExact_alpha_0.900}
	\end{subfigure}
\caption{
(\subref{subfig:Gamma_vs_beta_alpha_0.900}) The decay rate $\Gamma$ as a function of $\beta$ with fixed $\alpha = 0.90$, computed with four different methods: exact (the TDSE), NLO (the GY method), LO (naive dimensional analysis), and MC (the Monte Carlo method). (\subref{subfig:Gamma_ratio_vs_GammaExact_alpha_0.900}) The ratio $\Gamma / \Gamma_{\mathrm{exact}}$ versus $\Gamma_{\mathrm{exact}} / m$ as a reparametrization of the axes of (\subref{subfig:Gamma_vs_beta_alpha_0.900}) to facilitate comparisons. The solid line in (\subref{subfig:Gamma_ratio_vs_GammaExact_alpha_0.900}) is the constant $1$.}

\label{fig:vary_beta}
\end{figure}

\begin{figure}[!htb]
\centering
	\begin{subfigure}[ht]{0.48\textwidth}
		\centering
		\includegraphics[width=\textwidth]{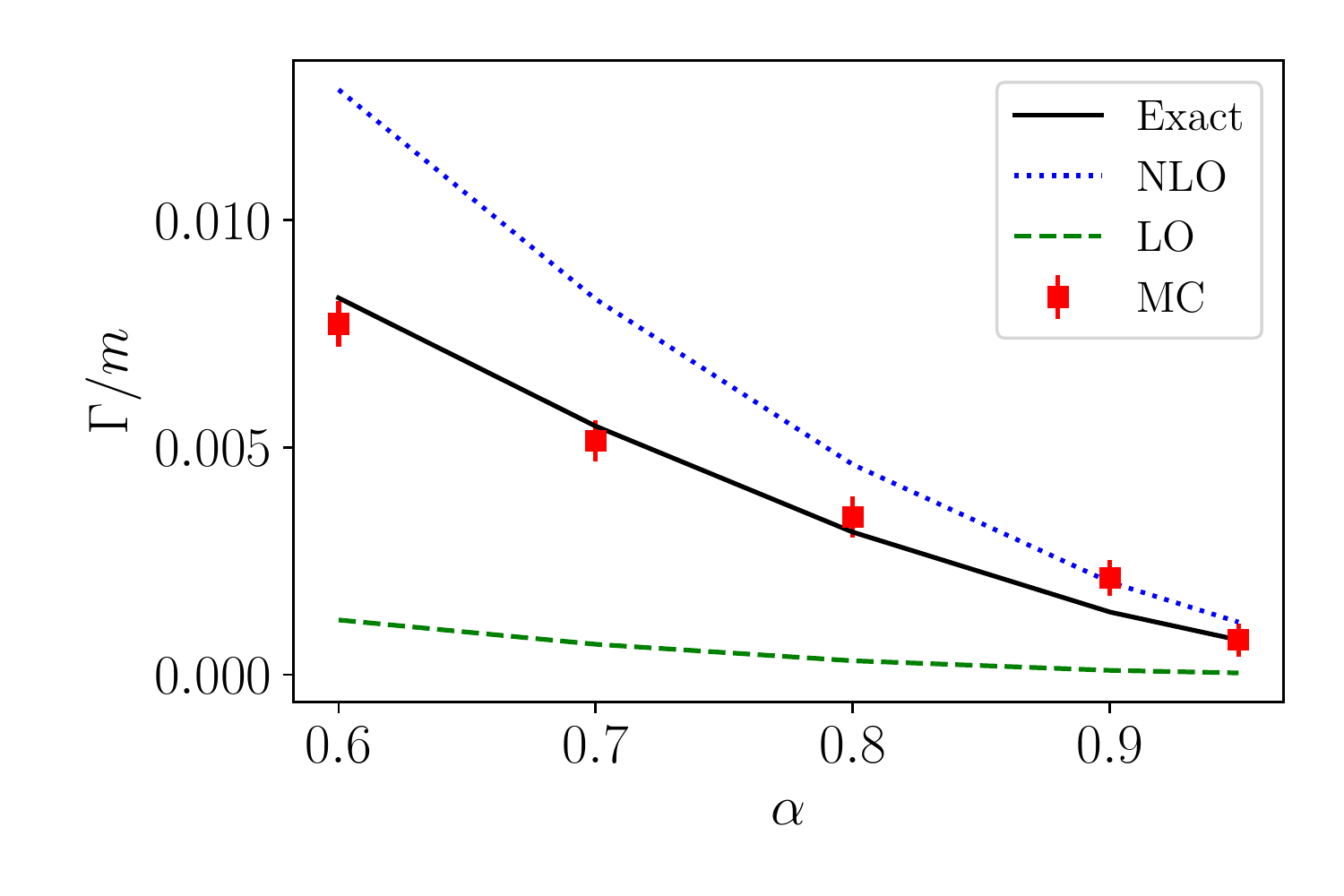}
		\caption{}
		\label{subfig:Gamma_vs_alpha_beta_9.000}
	\end{subfigure}
	\hfill
	\begin{subfigure}[!htbp]{0.48\textwidth}
		\centering
		\includegraphics[width=\textwidth]{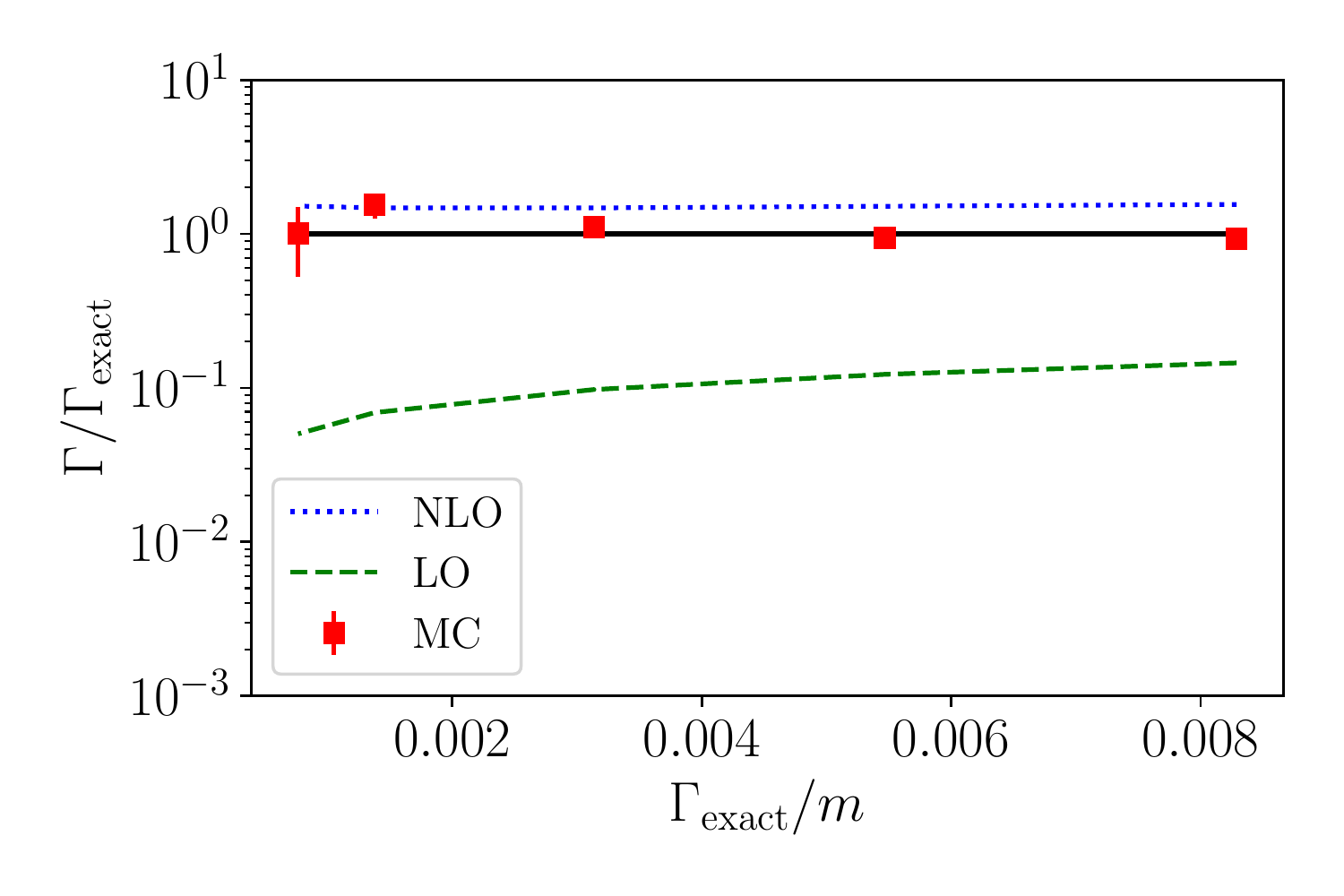}
		\caption{}
		\label{subfig:Gamma_ratio_vs_GammaExact_beta_9.000}
	\end{subfigure}
\caption{
(\subref{subfig:Gamma_vs_alpha_beta_9.000}) 
The decay rate $\Gamma$ as a function of $\alpha$ with fixed $\beta = 9.0$, computed with four different methods: exact (the TDSE), NLO (the GY method), LO (naive dimensional analysis), and MC (the Monte Carlo method). (\subref{subfig:Gamma_ratio_vs_GammaExact_beta_9.000}) The ratio $\Gamma / \Gamma_{\mathrm{exact}}$ versus $\Gamma_{\mathrm{exact}} / m$ as a reparametrization of the axes to facilitate comparisons. The solid line in (\subref{subfig:Gamma_ratio_vs_GammaExact_beta_9.000}) is the constant $1$.}
\label{fig:vary_alpha}
\end{figure}

\begin{figure}[!htb]
\centering
	\centering
	\begin{subfigure}[ht]{0.48\textwidth}
		\centering
		\includegraphics[width=\textwidth]{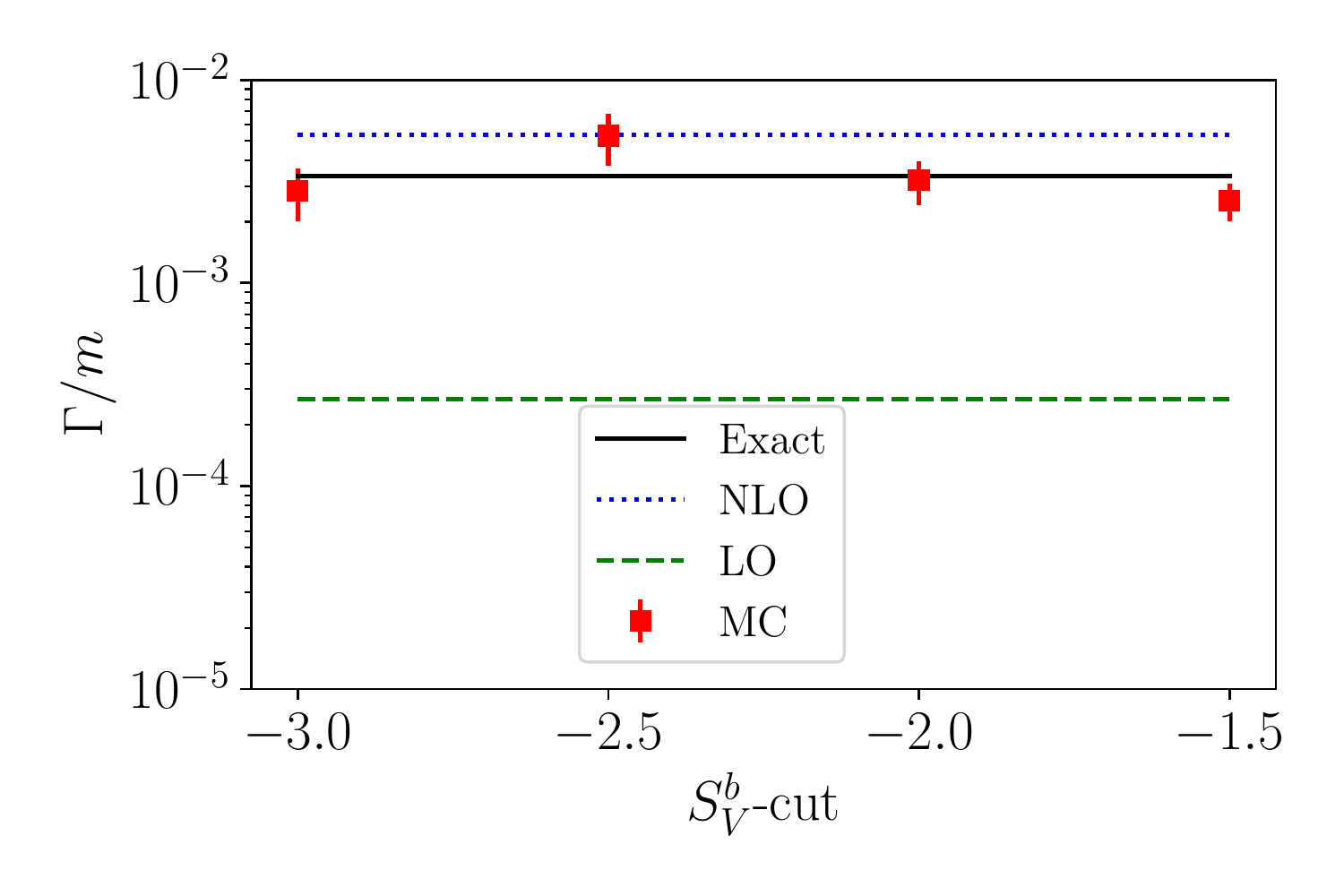}
		\caption{}
		\label{subfig:Gamma_vs_SbV-cut_alpha_0.900_beta_8.000}
    \end{subfigure}
    \hfill
	\begin{subfigure}[ht]{0.48\textwidth}
    	\centering
    	\includegraphics[width=\textwidth]{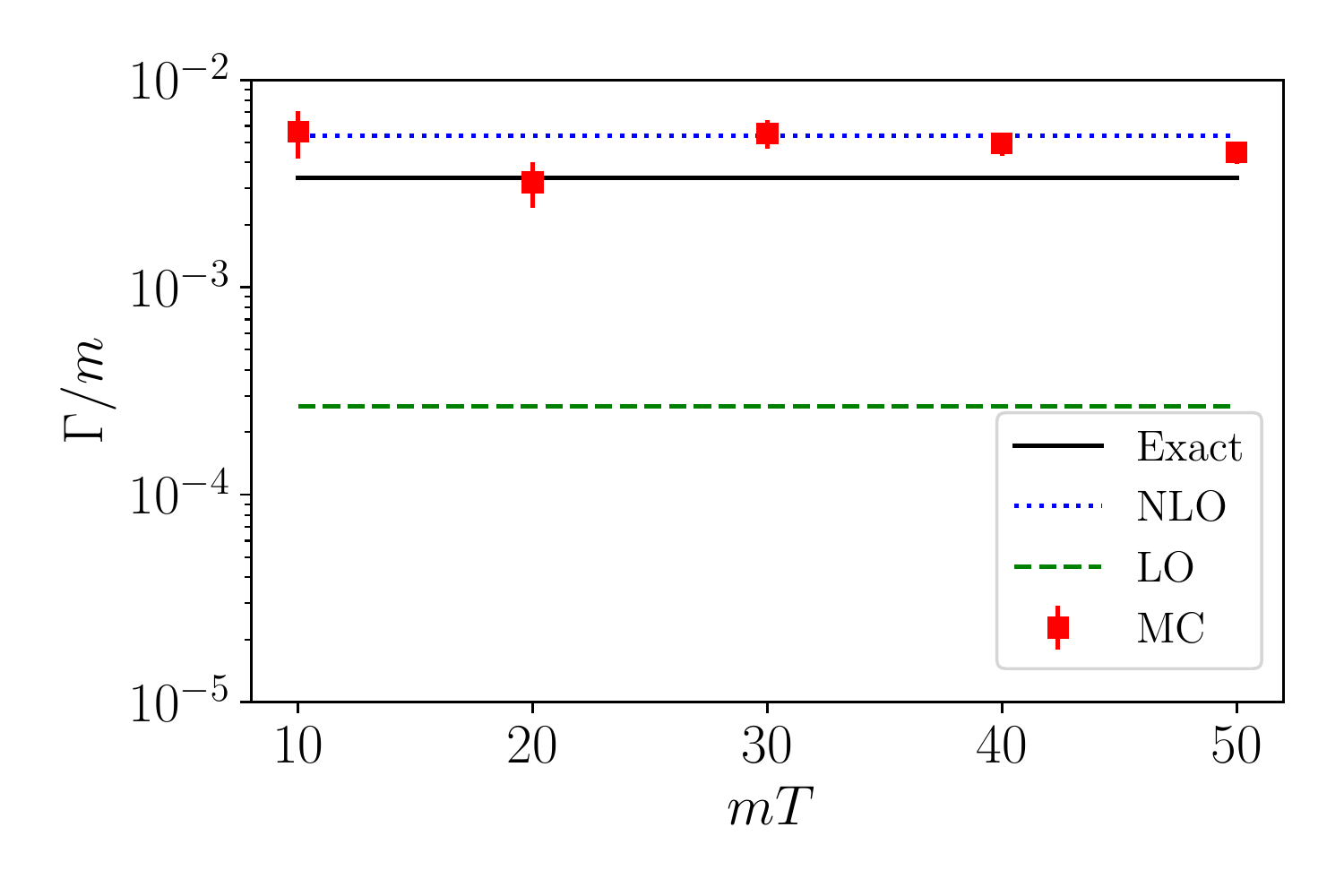}
    	\caption{}
    	\label{subfig:Gamma_vs_t_alpha_0.900_beta_8.000}
    \end{subfigure}
    \begin{subfigure}[ht]{0.48\textwidth}
    	\centering
    	\includegraphics[width=\textwidth]{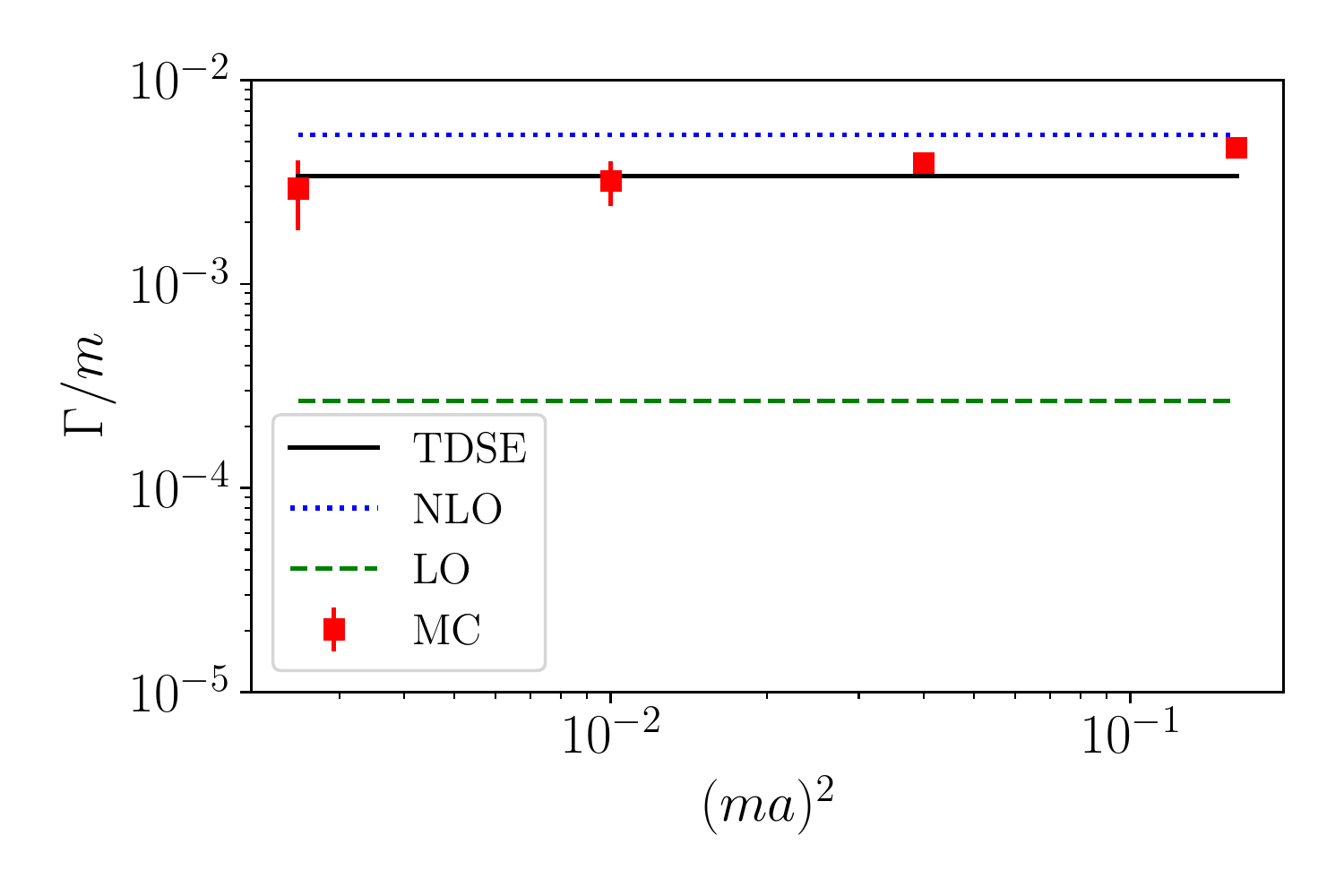}
    	\caption{}
    	\label{subfig:Gamma_vs_a_t_20.00_alpha_0.900_beta_8.000}
    \end{subfigure}
	\caption{Decay rate $\Gamma$  at $\beta = 8.0$ and $\alpha = 0.9$ computed with four different methods: exact (the TDSE), NLO (the GY method), LO (naive dimensional analysis), and MC (the Monte Carlo method). (\subref{subfig:Gamma_vs_SbV-cut_alpha_0.900_beta_8.000}) has varying $S^b_V$-cut, fixed at $m T = 20.0$ and $m a = 0.1$. (\subref{subfig:Gamma_vs_t_alpha_0.900_beta_8.000}) has varying $T$, fixed at $S^b_V\text{-cut} = -2.0$ and $m a = 0.1$. (\subref{subfig:Gamma_vs_a_t_20.00_alpha_0.900_beta_8.000}) has varying $a$, fixed at $S^b_V\text{-cut} = -2.0$ and $m T = 20.0$.}
	\label{fig:vary_SbV_cut_T_a}
\end{figure}

The parameters used in our MC ensembles are given in Table~\ref{table:ensembles} in Appendix~\ref{sec:details_computation}, and results are shown in Figs.~\ref{fig:vary_beta}--\ref{fig:vary_SbV_cut_T_a}, including both variation of model parameters ($\alpha$ and $\beta$; Figs.~\ref{fig:vary_beta} and~\ref{fig:vary_alpha} respectively) and variation of lattice/ensemble parameters [$S_V^b$-cut, $T$, and $a$; Figs.~\ref{fig:vary_SbV_cut_T_a}(\subref{subfig:Gamma_vs_SbV-cut_alpha_0.900_beta_8.000}),~~\ref{fig:vary_SbV_cut_T_a}(\subref{subfig:Gamma_vs_t_alpha_0.900_beta_8.000}), and~~\ref{fig:vary_SbV_cut_T_a}(\subref{subfig:Gamma_vs_a_t_20.00_alpha_0.900_beta_8.000}), respectively.]

Since we work in units where the mass is unity, and other scales in the problem like the spatial size of the semiclassical bounce solution are expected to be of this order,\footnote{In quantum field theory, the bounce can easily be much larger than the scalar mass parameter since it scales as the inverse of the semiclassical energy splitting between the true and false vacua. This is an effect of a friction term in the equation of motion defining the bounce. In quantum mechanics the friction term is not present, and to obtain a bounce much larger than the input mass scale requires an exponential tuning of the energy splitting. Typically, the bounce is still larger than $1/m$, so our estimate $\Gamma_{DA}=m e^{-S_b}$ is actually larger than the usual LO estimate $\Gamma_{\mathrm{LO}}=R_b^{-1}e^{-S_b}$ common in the literature. We see from the figures that the latter would only worsen the discrepancy of the LO estimate with the NLO, TDSE, and MC results.} we mostly work with lattice spacing $am=0.1$ and volume $N_T=400$. These choices are expected to avoid large corrections from lattice artifacts and finite volume effects, which we validate by varying these choices in two of the analyses described below. The ensemble-level $S_V^b$-cut is mostly taken to be $-2.0$, which is large enough in magnitude to avoid impacting the postselection  $S_V^b$-cut, while at the same time preventing the ensemble from probing configurations too close to the true vacuum, where it could get stuck. With these reasonable choices for the lattice/ensemble parameters, we compute $\Gamma$ for a range of potentials defined by $\alpha$ and $\beta$.

In Fig.~\ref{fig:vary_beta} we vary $\beta$ with fixed $\alpha$.  From the semiclassical perspective, varying $\beta$ is a probe of the LO exponential factor, $\Gamma\sim e^{-\beta \overline{S_b}}$. We find that the MC computation matches the exact TDSE result up to a factor $<2$ over a range $\Gamma/m \sim 10^{-4}\text{--}10^{-1}$. In the same range the NLO GY method achieves similar accuracy, with somewhat worse performance at higher rates. The LO estimate (DA) with dimensional analysis typically underestimates the rate by around an order of magnitude for these parameters.

In Fig.~\ref{fig:vary_alpha} we vary $\alpha$ at fixed $\beta$. From the semiclassical perspective, this is a probe of the mild $\alpha$-dependence of the LO exponential factor $e^{-\beta \overline{S_b}}$ (since $\overline{S_b}$ only depends on $\alpha$), as well as beyond-LO effects. Our MC results are in good agreement with both the TDSE and  GY results. They are closer to the ``exact" TDSE values than GY, which could be an indication that our MC method for computing $\Gamma$ is capable of accurately capturing some information beyond the NLO semiclassical approximation.

In Fig.~\ref{fig:vary_SbV_cut_T_a}(\subref{subfig:Gamma_vs_SbV-cut_alpha_0.900_beta_8.000}), with all other parameters fixed, we vary the ensemble-level $S_V^b$ cuts. The results from these ensembles are expected to be about the same. There is an uncertainty in finding the minimum of the probability distribution $p(S_V^b)$ measured on the ensemble, and this is the primary source of discrepancy among the values in Fig.~\ref{fig:vary_SbV_cut_T_a}(\subref{subfig:Gamma_vs_SbV-cut_alpha_0.900_beta_8.000}). In principle, the minimum should be nearly independent of the $S_V^b$-cut at ensemble generation, but there is an uncertainty introduced by numerical minimization with a finite sample. As  shown in Table~\ref{table:ensembles} in Appendix~\ref{sec:details_computation}, the postselection $S_V^b$-cuts for the these ensembles are not the same, although they are all around $-0.5$. The uncertainty in the postselection $S_V^b$-cuts is not reflected in the statistical error bars in Fig.~\ref{fig:vary_SbV_cut_T_a}(\subref{subfig:Gamma_vs_SbV-cut_alpha_0.900_beta_8.000}).

In Fig.~\ref{fig:vary_SbV_cut_T_a}(\subref{subfig:Gamma_vs_t_alpha_0.900_beta_8.000}), we vary $N_T$ with all other parameters fixed to test the $T$-dependence of our results. As  discussed in Sec.~\ref{sec:probabilitydensities}, we expect the $T$-dependence of the measured quantity $\rho_b(T)$ to be weak when $1/m \ll T \ll 1 / \Gamma$. With $a = 0.1 m^{-1}$, $N_T$ ranges from $200$ to $1000$, so $T = a N_T / 2$ ranges from $10 m^{-1}$ to $50 m^{-1}$. With $\alpha = 0.9$ and $\beta = 8.0$, the value of $1/\Gamma$ obtained by solving the TDSE is about $300 m^{-1}$, so the condition $1/m \ll T \ll 1 / \Gamma$ is satisfied. There is some mild variation in the MC results as we vary $T$, but within statistical uncertainties they fall between the TDSE and GY results for this model point, and the uncertainty in $\Gamma$ associated with residual $T$-dependence is again a factor $<2$.

Finally, in Fig.~\ref{fig:vary_SbV_cut_T_a}(\subref{subfig:Gamma_vs_a_t_20.00_alpha_0.900_beta_8.000}) we vary the lattice spacing $a$ with other parameters fixed. There is an $O(a^2)$ difference between the lattice action and the continuum action, so reducing the value of $a$ can make the result more precise. Since $m$ is the characteristic scale in the continuum theory, $a$ should not be substantially greater than $1/m$. However, for fixed time range $T$, smaller $a$ leads to a greater number of sites $N_T = 2 T / a$, and greater computational cost. We find that values of $a$ in the range $[0.05 m^{-1}, 0.4 m^{-1}]$ all give accurate results, justifying the use of $a = 0.1 m^{-1}$ for the majority of our previous computations.

\section{Long Lifetimes\label{sec:longlifetime}}

In the previous section,  we saw that straightforward ensemble generation with a hard wall on the quantity $S_b^V$ allows an accurate computation of the probability density $\rho$ and thus a good estimate of the decay rate, when these quantities are not too small. However, when the lifetime becomes very long, direct generation of the ensembles becomes impractical: starting from the vicinity of the false vacuum, the saddle point is simply too difficult to find by random fluctuations.

Instead, we consider a modification of the computation which we refer to as {\emph{constrained ensemble reweighting}}. In the ensemble generation, we fix the trajectories to the classical turning point $b$ at the midpoint in Euclidean time. In doing so we give up time translation invariance and the associated improvement in statistics, but we gain much more by ``telling" the MC that it needs to reach $b$. 
To be more precise, for each rate computation, we generate two ensembles, one with the constraint applied and one without, and attempt to compute the probability of finding configurations from the constrained ensemble in the unconstrained ensemble.

In an ensemble of $N$ configurations with $N_T$ sites, the number of configurations $\Delta N[x^{\star}]$ near a given configuration $\{x^{\star}_i\}_{1 \leq i \leq N_t}$ in a vicinity of volume $\prod_{i=1}^{N_T} \Delta x_i$ is given by
\begin{align}
    \frac{1}{N} \frac{\Delta N \bracket{x^{\star}}}{\prod_{i=1}^{N_T}\Delta x_i} \approx c e^{- S \bracket{x^{\star}}}
    ,
\label{eq:prob_vicinity}
\end{align}
where $c$ is a normalization factor.
The ensemble generation may have some imposed constraints in the space of configurations. These constraints affect which configurations are allowed but still retain the relative probabilities of allowed configurations. The factor $c$ may depend on the constraints but does not depend on configurations $x^{\star}$ as long as $x^{\star}$ is not forbidden by the constraints. $c$ is also independent of the total number of configurations $N$.

For an ensemble with $N_1$ configurations generated by the modified double-well potential we are interested in, which we denoted as ``ensemble 1,'' we first consider $x^{\star}\equiv x^{\star}_i = x_{\mathrm{FV}}$ for all $i$, \textit{i.e.}, the FV-static configuration. The number of configurations in the vicinity of the static $x_{\mathrm{FV}}$ configuration is given by
\begin{align}
    \frac{1}{N_1} \frac{\Delta N_1 \bracket{x_{\mathrm{FV}}}}{\prod_{i=1}^{N_T}\Delta x_i} \approx c_1 e^{- S\bracket{x_{\mathrm{FV}}}}
    .
\label{eq:N1_FV}
\end{align}
Now consider $x^{\star} = x_b$, the bounce solution, in the same ensemble. Configurations in its vicinity are representative contributors to $\rho$. The number of such configurations is
\begin{align}
    \frac{1}{N_1} \frac{\Delta N_1 \bracket{x_b}}{\prod_{i=1}^{N_T}\Delta x_i} \approx c_1 e^{- S\bracket{x_b}}
    .
\label{eq:N1_b}
\end{align}
Therefore, with the same volume $\prod_{i=1}^{N_T}\Delta x_i$, $\Delta N_1 [x_b] / \Delta N_1 [x_{\mathrm{FV}}] \approx \exp(- S[x_b] + S[x_{\mathrm{FV}}])$ is exponentially suppressed in the semiclassical limit. In such a case, from ensemble 1, $\Delta N_1 [x_{\mathrm{FV}}]$ is measurable whereas $\Delta N_1 [x_b]$ is difficult to measure.

To circumvent the exponential suppression we can generate a second ensemble, denoted as ``ensemble 2,'' with $N_2$ configurations constrained by $x_{N_T / 2} = b$, corresponding to a center time constraint $x(t=0) = b$ in the continuum. Due to this constraint, effectively there are  now only $N_T - 1$ sites on the lattice. The number of configurations in ensemble 2 in the vicinity of a configuration $x^{\diamond}$  is
\begin{align}
    \frac{1}{N_2} \frac{\Delta N_2 \bracket{x^{\diamond}}}{\prod_{1 \leq i \leq N_T,\, i \neq N_T / 2}\Delta x_i} \approx c_2 e^{- S\bracket{x^{\diamond}}}
    ,
\label{eq:N2_b}
\end{align}
where $c_2$ is a normalization factor different from $c_1$ (and even has a different dimension, $[c_2] = [x] [c_1]$).  In ensemble 2, false-vacuum-like configurations are not allowed due to the constraint, so for relevant configurations near the bounce, $x^{\diamond}\sim x_b$, $\Delta N_2 \bracket{x_b}$ is numerically calculable without suffering from an exponential suppression. 

We can use Eqs.~(\ref{eq:N1_FV}) and~(\ref{eq:N2_b}) to estimate the probability density at $x = b$. We write
\begin{equation}
\begin{aligned}
     \rho_b 
    \approx & \frac{1}{N_1} \left.\frac{d N_1 }{d x_{N_T/2}}\right|_{x_{N_T/2} = b} \\
    = & \frac{1}{N_1} \paren{\prod_{i \neq N_T / 2} \int d x_i\,} \left. \frac{d^{N_T} N_1 \bracket{x_1, \cdots \cdots, x_{N_T}}}{\paren{\prod_{i} d x_i}}\right|_{x_{N_T/2} = b} \\
    = & \paren{\prod_{i \neq N_T / 2} \int d x_i\,} c_1 e^{-S\bracket{x}} \\
    = & \frac{c_1}{c_2} \paren{\prod_{i \neq N_T / 2} \int d x_i\,} c_2 e^{-S\bracket{x}} \\
    = & \frac{c_1}{c_2} \paren{\prod_{i \neq N_T / 2} \int d x_i\,} \frac{1}{N_2} \frac{d^{N_T - 1} N_2 \bracket{x_1, \cdots, x_{N_T/2 - 1}, x_{N_T/2 + 1}, \cdots, x_{N_T}; \, x_{N_T/2} \approx b}}{\prod_{i \neq N_T/2} d x_i} \\
    = & \frac{c_1}{c_2}. \\
\end{aligned}
\end{equation}
Thus we extract the decay rate,
\begin{align}
    \Gamma/\sqrt{m}\approx \rho_b\approx c_1/c_2
    \label{eq:gammac1c2}
\end{align}
where $c_1/c_2$ can be computed from two ensembles as
\begin{equation}
    \frac{c_1}{c_2} \approx e^{- S\bracket{x^{\diamond}} + S\bracket{x_{\mathrm{FV}}}} \left. \paren{\frac{1}{N_1} \frac{\Delta N_1 \bracket{x_{\mathrm{FV}}}}{\prod_{i=1}^{N_T}\Delta x_i}} \middle/ \paren{\frac{1}{N_2} \frac{\Delta N_2 \bracket{x^{\diamond}}}{\prod_{1 \leq i \leq N_T,\, i \neq N_T / 2}\Delta x_i}} \right. .
\label{eq:c1_over_c2}
\end{equation}

There is still an ``exponentially hard'' aspect of the method: for large lattices the probability of finding a configuration in a volume $\prod_{i=1}^{N_T}\Delta x_i$ near another configuration is exponentially small in $N_T$. To ameliorate this we find that it is sufficient to work with somewhat larger lattice spacings and smaller volumes, without substantially sacrificing accuracy.

We test the method on a benchmark point with $\alpha = 0.9$, $\beta = 60.0$, and  we generate  two ensembles  with $a = 0.3 m^{-1}$, $N_T = 120$. As described above, in ensemble 2 we impose a constraint $x (t = 0) = b$ and $S_V^b = -1.20$ during the ensemble generation to avoid the dominance of true-vacuum-like configurations. The number of configurations in ensemble 1 is $N_{\mathrm{cf}, 1} = 100,000$. Ensemble 1 has no constraint at $x (t = 0)$, and we have effectively set no $S_V^b$-cut either, because $\beta$ is very large. It is highly improbable for a configuration in ensemble 1 to approach $b$, by a factor of order $e^{- \beta \overline{S_b}} \approx 10^{-27}$, and we find that all configurations have $S_V^b = 0$. Therefore in the formulas above $N_1 = N_{\mathrm{cf}, 1} = 100,000$. 

For ensemble 2 we still need to impose cuts on $S_V^b$, similar to the procedure described in Sec.~\ref{sec:numerics}. With the additional constraint $x (t = 0) = b$, the detailed arguments provided in Appendix~\ref{sec:sbv}, used to justify the particular postselection cut on $S_V^b$ used in  Sec.~\ref{sec:numerics}, do not hold exactly. However, the general principle that the cut should be chosen prior to the onset of the exponential rise in the $S_V^b$ distribution still applies, and in practice we find that the same choice of postselection cut $S_V^b>-0.5$ is adequate. In general the variation of the cut within a range that does not sample the exponential rise, or approach unnecessarily close to zero, leads to an  $O(1)$ impact  on the final result for the rate. This would be a reasonable  target accuracy for this method, but in our initial investigation here we find somewhat larger sources of error. After postselection, ensemble 2 contains $N_2=48,674$ configurations.  

\begin{figure}[!t]
\centering
		\includegraphics[width=0.65\textwidth]{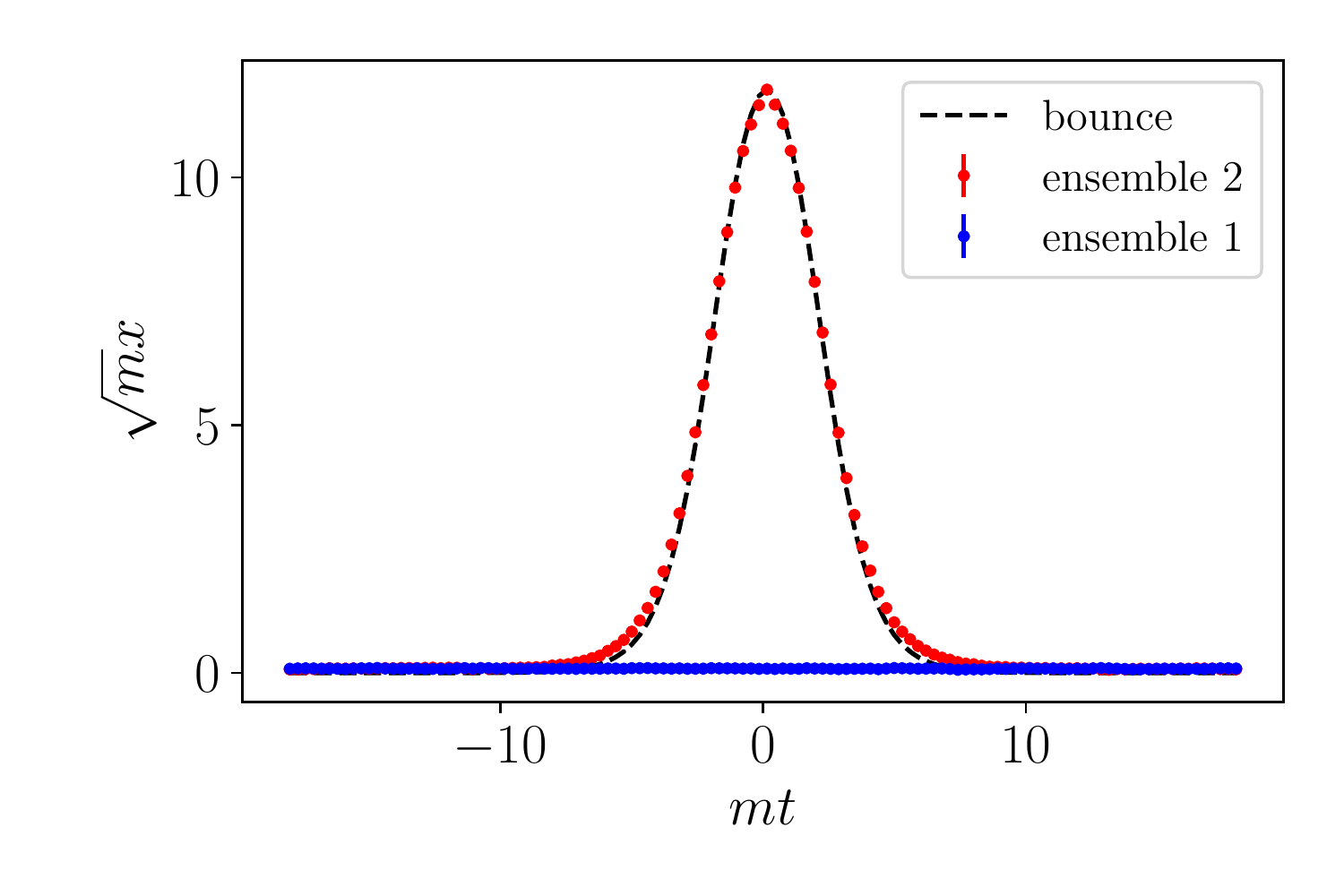}
\caption{Median configurations for ensembles 1 and 2. The numerical bounce solution obtained from solving the classical equation of motion is shown for comparison. Statistical error bars on the median configurations are quite small, with maximum errors in $x$ of $0.00053 m^{-1/2}$ for ensemble 1 and $0.0025 m^{-1/2}$ for ensemble 2. These uncertainties are tiny compared to the characteristic scale of variation of the potential. Because other errors are much larger, we  neglect this source of uncertainty in subsequent error estimates.}

\label{fig:central_config}
\end{figure}

To carry out the analysis 
we must define the configurations around which to count neighboring configurations in each ensemble. For ensemble 1 we could simply use $x^{\star}=x_{\mathrm{FV}}(t) = 0$, as used in the formulas above. For ensemble 2, a convenient choice for $x^{\diamond}$ is to construct a smoothed configuration by taking the mean or median value of $x$ evaluated at each $t$ over all the configurations in the postselected ensemble.  We use the median configuration, shown in Fig.~\ref{fig:central_config}, to reduce the effects of possible outliers, but the mean configuration is in fact extremely similar. (To keep the ensembles on the same footing, we also use the median configuration in ensemble 1 for $x^{\star}$ rather than directly using $x^{\star} = x_{\mathrm{FV}} = 0$, but the difference is negligible and we continue to refer to the central configuration for this ensemble as $x_{\mathrm{FV}}$.) We also overlay the semiclassical bounce solution in Fig.~\ref{fig:central_config}, demonstrating, as a by-product, that the smoothed configurations  closely approximate the bounce, as one might expect deep in the semiclassical regime.

The vicinity of the median configuration is defined by choosing the windows $\{\Delta x_i\}$. In principle we would like all $\Delta x_i$  to be infinitesimal, but this is not possible in practice, because the number of configurations in the neighborhood is exponentially small in the number of sites $N_T$. Instead, we take $\Delta x_i$ to be finite at order $O(\sqrt{\beta / m})$, \textit{i.e.}, the characteristic scale of the potential in $x$-space. For simplicity, we choose $\Delta x_i\equiv \Delta x$ to be site-independent. For ensemble 2 a configuration $x(t_i)$ is identified as lying in the vicinity of $x^{\diamond}$ if $x^{\diamond}(t_i) - \Delta x / 2 < x(t_i) < x^{\diamond}(t_i) + \Delta x / 2$ for all sites $i$, and similarly for ensemble 1. We check this criterion for all configurations after postselection, and the number of configurations that pass the test give the values of $\Delta N_2[x^{\diamond}]$ and $\Delta N_1[x_{\mathrm{FV}}]$ in Eq.~(\ref{eq:c1_over_c2}).

In Eq.~(\ref{eq:c1_over_c2}), there is the factor $e^{- S\bracket{x^{\diamond}} + S\bracket{x_{\mathrm{FV}}}}$ which can be computed from the median-smoothed configurations in each ensemble. However, when the vicinity defined by $\Delta x$ is finite, the action of every actual configuration in the neighborhood receives large contributions from high frequency fluctuations. Therefore we consider a second method to estimate the difference  $- S\bracket{x^{\diamond}} + S\bracket{x_{\mathrm{FV}}}$. We construct the sample distribution of the action over each neighborhood of original configurations and identify the action difference with the difference in the means of these distributions. The distributions are peaked at much higher values of $S$ than the action of the median smoothed configurations, due to the high-frequency fluctuations in the original configurations~(see Appendix~\ref{sec:num_details_longlifetime} for numerical details). Loosely speaking we can think of this alternate prescription as redefining the central configuration by a typical configuration in the neighborhood of the smoothed one.

\begin{figure}[!htb]
\centering
	\begin{subfigure}[ht]{0.48\textwidth}
		\centering
		\includegraphics[width=\textwidth]{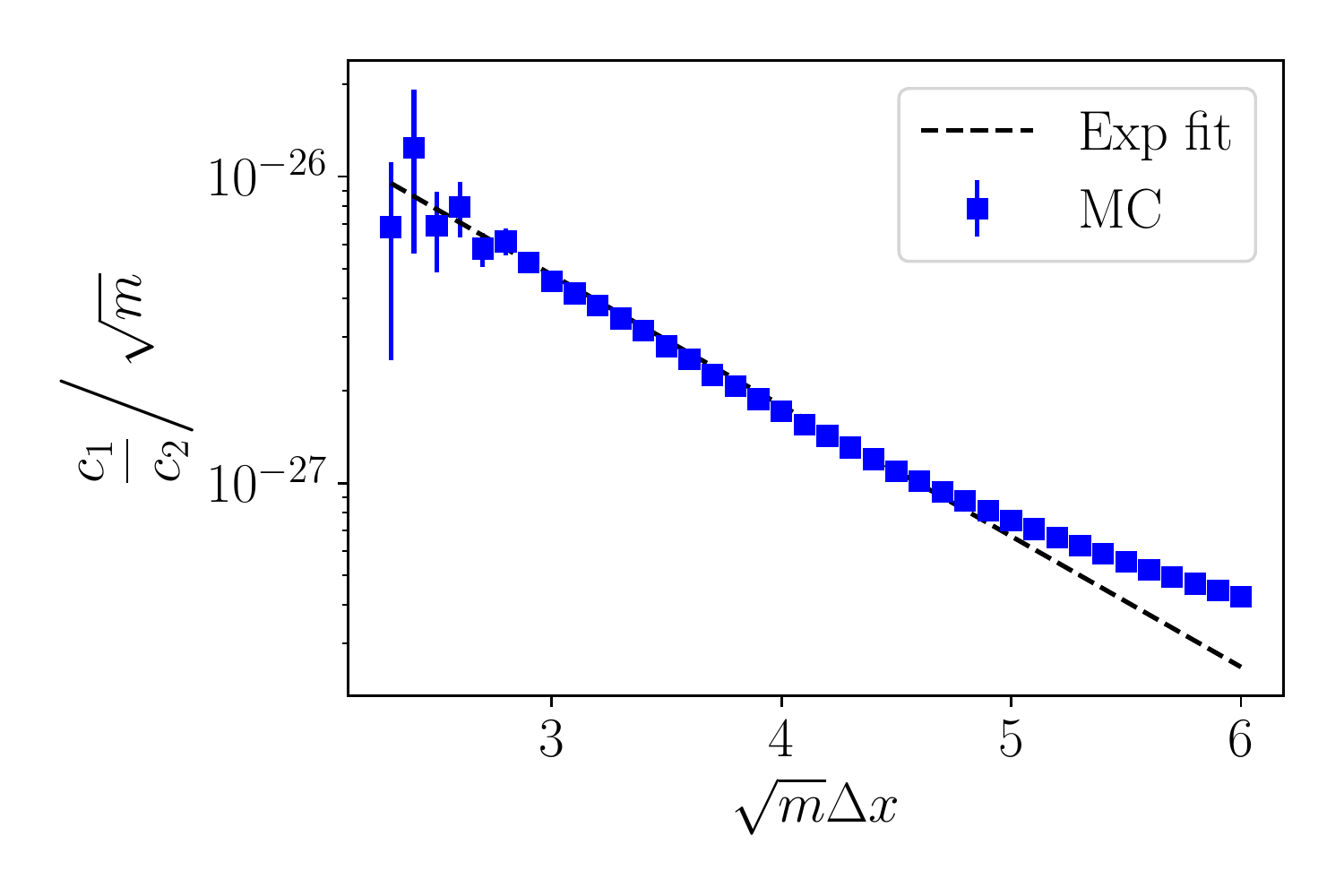}
		\caption{$e^{- S\bracket{x^{\diamond}} + S\bracket{x_{\mathrm{FV}}}}$ is computed directly from the median-smoothed configurations.}
		\label{subfig:c1_over_c2_panel_1}
	\end{subfigure}
	\hfill
	\begin{subfigure}[!htbp]{0.48\textwidth}
		\centering
		\includegraphics[width=\textwidth]{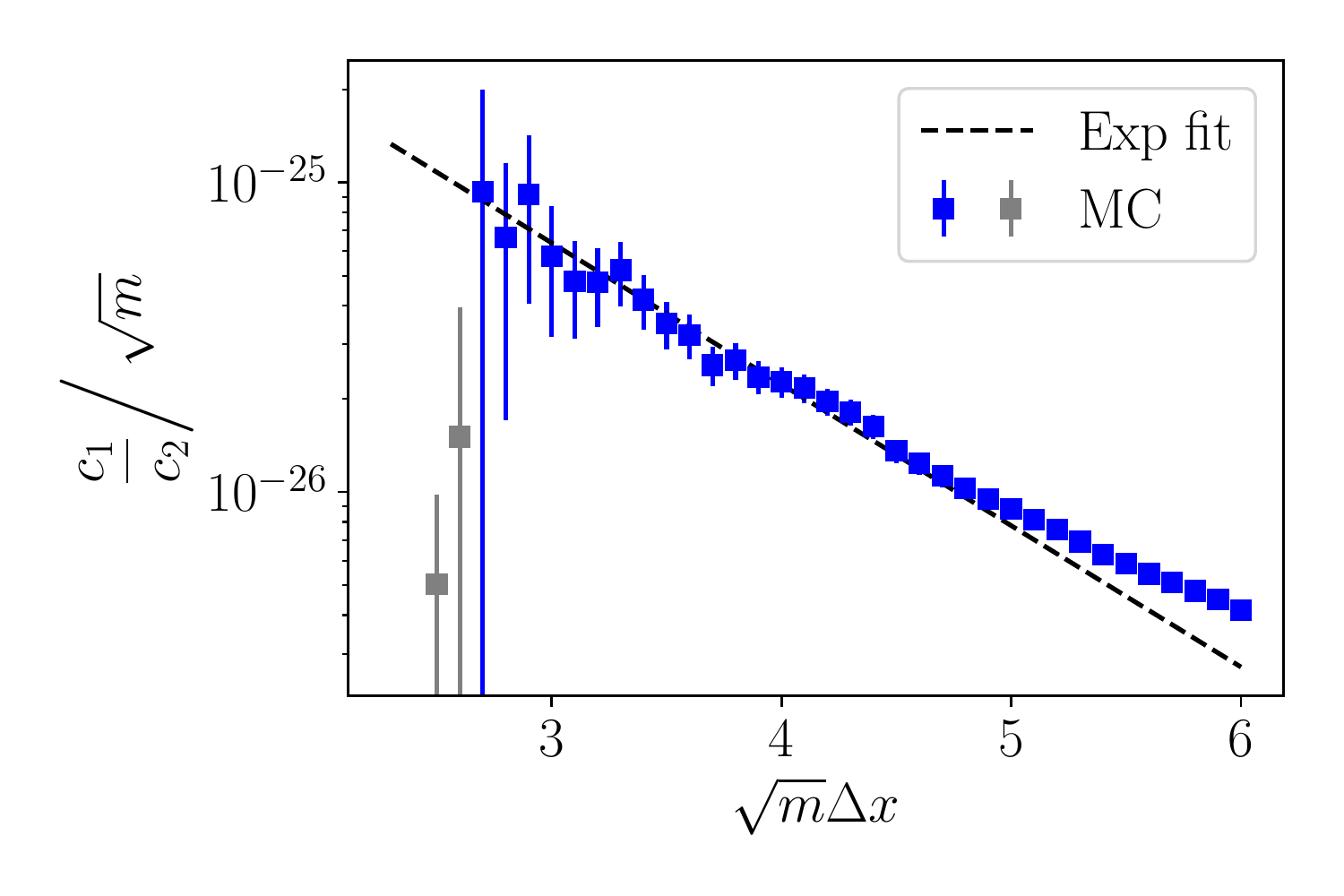}
		\caption{$e^{- S\bracket{x^{\diamond}} + S\bracket{x_{\mathrm{FV}}}}$ is computed using the means of the sample distributions of the actions of configurations near $x_{\mathrm{FV}}$ and $x^{\diamond}$. Data marked by gray points are not used in the fit.}
		\label{subfig:c1_over_c2_panel_2}
	\end{subfigure}
\caption{Values of $c_1 / c_2$ obtained using the two different approaches to compute  $e^{- S\bracket{x^{\diamond}} + S\bracket{x_{\mathrm{FV}}}}$ as described in the text. The dashed line shows an exponential fit, which we extrapolate to zero to obtain  estimates for $\Gamma$. We use these results together with the unextrapolated values near the smallest accessible $\Delta x$ to obtain a conservative uncertainty range $\Gamma = 10^{-25\pm1}m$. The central value is quite close to the semiclassical NLO estimate  $\Gamma_{\mathrm{GY}} = 8.61 \times 10^{-26} m$, while the leading order estimate is about 2 orders of magnitude smaller, $\Gamma_{\mathrm{DA}} = 1.56 \times 10^{-27} m$.
}
\label{fig:c1_over_c2}
\end{figure}

In Fig.~\ref{fig:c1_over_c2}, we use these two different prescriptions for the exponential factor in~(\ref{eq:c1_over_c2}) to compute $ c_1 / c_2$ with finite-sized neighborhoods. The statistical uncertainties are greater at smaller $\Delta x$ because fewer configurations survive. At intermediate $\Delta x$ the results are very close to an exponential function of $\Delta x$. Heuristically this can be understood as follows. In ensemble 1, fluctuations can only raise the action, so as $\Delta x$ increases the number of neighboring configurations rapidly saturates to an $\calO(1)$ fraction of the total in  ensemble 1.  In ensemble 2 the fluctuations do not necessarily raise the action and saturation only occurs at larger $\Delta x$. These behaviors are reflected in Fig.~\ref{fig:c1_over_c2}. The difference in the typical action of fluctuations then implies an exponential difference in the $\Delta x$ distribution of configurations which is measured by $c_1/c_2$. By contrast, even in the second method, the exponential prefactor is highly stable with $\Delta x$, as shown in Fig.~\ref{fig:c1_over_c2}. (In the first method this factor does not change, by definition.) 

We now perform four estimates of the decay rate from these results, corresponding to each of the two methods of computing the difference  $- S\bracket{x^{\diamond}} + S\bracket{x_{\mathrm{FV}}}$, and taking the results in Fig.~\ref{fig:c1_over_c2} with and without exponential extrapolation to $\Delta x=0$. 
With exponential extrapolation, we fit $c_1 / c_2$ as a function of $\Delta x$ to the form $p_0 \exp(-p_1 \Delta x)$ where $p_0$ and $p_1$ are fit parameters. Since we are only interested in semiquantitative extrapolation, we use a naive fit that ignores the correlation among data at different $\Delta x$ and treats them as uncorrelated. 
In this way we obtain a conservative estimate of the uncertainties arising from practical limitations on the smallest $\Delta x$ that can be accessed directly.

The values of $c_1 / c_2$ at different $\Delta x$ are correlated, and we perform exponential fits only using data with relatively small statistical uncertainties. The extrapolated results at $\Delta x = 0$ are (\ref{subfig:c1_over_c2_panel_1}) $c_1 / c_2 \approx 9.0 \times 10^{-26} \sqrt{m}$ and (\ref{subfig:c1_over_c2_panel_2}) $c_1 / c_2 \approx 1.5 \times 10^{-24} \sqrt{m}$. Since $c_1 / c_2$ is an estimate for $\rho_b \approx \Gamma / \sqrt{m}$, the results translate to (\ref{subfig:c1_over_c2_panel_1}) $\Gamma \approx 9.0 \times 10^{-26} m$ and (\ref{subfig:c1_over_c2_panel_2}) $\Gamma \approx 1.5 \times 10^{-24} m$. Without extrapolation, the values are of order $\Gamma\approx 10^{-26}m$ and $\Gamma\approx 10^{-25}m$ at the smallest $\Delta x$ with controlled statistical errors in the two methods. Putting the four results together we obtain
\begin{align}
     \Gamma \approx \paren{10^{-26} \textrm{--} 10^{-24}}m
    \label{eq:smallgammaresult}
\end{align}
with order-of-magnitude uncertainty associated with finite $\Delta x$. 

The semiclassical NLO estimate for the decay rate is $\Gamma_{\mathrm{GY}} = 8.61 \times 10^{-26} m$, while the leading order estimate is about two orders of magnitude smaller, $\Gamma_{\mathrm{DA}} = 1.56 \times 10^{-27} m$. The central value in Eq.~(\ref{eq:smallgammaresult}) is close to the NLO result and the conservative uncertainty band is still tighter than the LO-NLO difference.

We regard the method and analysis presented in this section as a promising first exploration of simple reweighting techniques for systems with long lifetimes.  To better control the uncertainties, a more rigorous argument for the exponential extrapolation is essential, and the two estimates of $- S\bracket{x^{\diamond}} + S\bracket{x_{\mathrm{FV}}}$ can be compared with larger ensembles across a range of potentials. Nonuniform $\Delta x_i$ might also provide a useful tool. We leave these directions to future work.

\section{Conclusions and Outlook}
\label{sec:concl}

In this work, we develop a new framework for studying systems with metastable vacua in Euclidean Monte Carlo simulations. Our main results are
\begin{enumerate}[(i)]
    \item In quantum mechanics with a metastable vacuum state in the potential, the decay rate can be estimated if the probability density is known at the classical turning point, as shown in Eq.~(\ref{eq:gammaappx}). The probability density can be expressed in terms of a lattice observable $\hat\rho$, see Eqs.~(\ref{eq:rhoequalsrhohat}), (\ref{eq:fdef}), (\ref{eq:rhohatdef}), and (\ref{eqAdef}).
    \item Direct lattice simulation is feasible if the lifetime is not too long and a wall is inserted to prevent the ensemble from wandering into the basin of the true vacuum. For this purpose we find that a cut on the total contribution to the potential energy from the classically allowed region, $S_V^b$, provides an effective barrier, Eq.~(\ref{eq:SbVdef}). We place a loose cut during ensemble generation and a tighter cut in postselection. A good choice for the latter is the minimum of the sample distribution of $S_V^b$. This cut avoids the need for any analytic continuation, while introducing an uncertainty into the final result.
    \item Testing the method over a family of example models, we find that we can reproduce the results of numerical exact diagonalization to similar or better accuracy than next-to-leading-order semiclassical analysis with the NLO prefactor computed numerically using the Gel'fand-Yaglom method. The differences are generally an $O(1)$ factor, while the leading order semiclassical estimate with prefactor fixed on dimensional grounds is generally off by more than an order of magnitude. The lattice results show satisfactory stability when varying over a range of lattice simulation parameters.
    \item For long lifetimes, a direct lattice computation is again infeasible, but we find that a simple modification of the technique is effective to compute the probability density at the classical turning point $b$: we generate an additional ensemble with a constraint that the trajectories reach $b$ at a fixed time. By a reweighting procedure we can then estimate $\rho(b)$ using Eqs.~(\ref{eq:gammac1c2}) and (\ref{eq:c1_over_c2}). In an example case this method gives results consistent with NLO semiclassics within an order of magnitude, while LO semiclassics differs by 2 orders of magnitude. The uncertainties are driven by an extrapolation and might be improved by refinements of the method.
\end{enumerate}
Our work is of an exploratory nature and as such we focus here on the simplest one-particle quantum mechanical theories. In these theories there are multiple other accurate means of computation (exact diagonalization, NLO semiclassics), which we use to benchmark our method.
Lattice techniques would be of limited interest if they were confined to one-particle quantum mechanics. 
Fortunately, there are reasons to be optimistic about 
the future 
extensions to multiparticle quantum mechanics and field theories. 
The main new aspects in the more complex theories are the presence of a classical turning surface, rather than a turning point, and of renormalization effects. A natural first step would be to generalize the probability density as a function of particle coordinate $x$ to a probability density in the energy of field configurations on spatial slices; the density at the turning point $b$ should then be replaced by the probability density at energy equal to that of the false vacuum. This energy is shifted by quantum effects, as are the model parameters in the usual way, and one could attempt to account for renormalization effects by standard lattice methods. Our analysis in Sec.~\ref{sec:theory} would need to be extended to obtain the relationship between $\Gamma$ and $\rho(E)$ appropriate for field theories. We hope to address this problem in future work.

Following the real-time evolution of metastable states is also an important problem for the nascent field of quantum simulations applied to high energy physics. It would be interesting to explore hybrid classical-quantum techniques utilizing the lattice methods developed here.

The most exciting application of lattice Monte Carlo techniques to theories with metastable vacua is in cases where a precise semiclassical formulation is not well-understood. These include scalar theories where the false vacua are not present in the classical potential, but are generated by quantum effects, and gauge theories where long-lived false vacua are believed to be generated by strong dynamics (e.g. Yang-Mills at large $N$~\cite{Witten:1998uka}.) Our work is only a first step in this direction, and both theoretical and computational developments are needed to perform accurate computations in all of the theories of interest.
It would be interesting to explore application of the multicanonical method~\cite{Berg:1991cf,Berg:1992qua,Moore:2000jw,Moore:2001vf,PhysRevD.106.114507}, which has been developed to address critical slowing down in systems with first-order phase transitions, to the case at hand with exponentially slow quantum tunneling.
In addition to the theoretical aspects mentioned above, on the computational side, smarter sampling such as creating ensembles using machine learning techniques~\cite{Kanwar:2020xzo,Boyda:2020hsi} might improve the accuracy when the decay rates are very slow.
However, for the purpose of simply verifying the existence of metastable states, straightforward lattice simulations may in fact be quite effective.

\section*{Acknowledgments}
We thank Bhairav Valera for collaboration in the early stages of this work, and Di Luo, Bryan Clark, Oliver Gould, and Norikazu Yamada for useful discussions. 
This work was supported in part by the U.S.\ Department of Energy, Office of Science, 
Office of High Energy Physics under Award No. DE-SC0015655 and by its QuantISED program under
a grant for the Fermilab Theory Consortium ``Intersections of QIS and Theoretical Particle
Physics.'' 
A.\ El-Khadra was also supported in part by the Simons Foundation under its Simons Fellows in Theoretical Physics program. 
Computations for this work were carried out in part on facilities of the USQCD Collaboration, which are funded by the Office of Science of the U.S.\ Department of Energy. 

\begin{appendices}

\section{Decay rates from the Gel'fand-Yaglom method}

The NLO decay rate from the saddle point approximation is~\cite{Coleman:1977py} 
\begin{equation}
	\Gamma = \paren{\frac{S [x_b]}{2 \pi}}^{1/2} e^{- S [x_b]} \mathrm{Im} \sqrt{ \frac{ \det \bracket{ S''[x_\mathrm{FV}] } }{ \det' \bracket{ S''[x_b] } } }
	,
\end{equation}
where $\det '$ means the zero eigenvalue is removed from the determinant.
The two differential operators are
\begin{equation}
	S''[x_\mathrm{FV}] = - \frac{d^2}{d r^2} + V''(x_\mathrm{FV}(r))
	,
\end{equation}
\begin{equation}
	S''[x_b] = - \frac{d^2}{d r^2} + V''(x_b (r))
\end{equation}
where $r \equiv \abs{t}$ is the distance in Euclidean time from the center of the bounce.

It is more convenient to work with dimensionless quantities. The decay rate is then
\begin{equation}
	\Gamma = m \sqrt{\beta} \paren{\frac{\overline{S_b}}{2 \pi}}^{1/2} e^{-\beta \overline{S_b}} \mathrm{Im} \sqrt{ \frac{ \det \bracket{ \bar{S}''[\bar{x}_\mathrm{FV}] } }{ \det' \bracket{ \bar{S}''[\bar{x}_b] } } }
	.
\label{eq:Gamma_GY_dimensionless}
\end{equation}
The potential in the dimensionless form is
\begin{equation}
    \bar{V} \paren{\bar{x}} =  \begin{cases} 
      \frac{1}{2} \bar{x} - \frac{1}{2} \bar{x}^3 + \frac{\alpha}{8} \bar{x}^4 & \bar{x} < \bar{x}_{\mathrm{TV}} \\
      \bar{V}_{\mathrm{TV}} & \bar{x} \geq \bar{x}_{\mathrm{TV}}
      .
   \end{cases}
\end{equation}
In the semiclassical limit, the flat region at $\bar{x} \geq \bar{x}_{\mathrm{TV}}$ does not affect the result, and we can instead use $\bar{V} \paren{\bar{x}} = \frac{1}{2} \bar{x} - \frac{1}{2} \bar{x}^3 + \frac{\alpha}{8} \bar{x}^4$ for $\bar{x} \in \mathbb{R}$.
We denote
\begin{equation}
    \mathcal{M} \equiv \bar{S}''[\bar{x}_b] = - \frac{d^2}{d \bar{r}^2} + 1 + \mathcal{V} \paren{\bar{r}} ,
\end{equation}
\begin{equation}
    \mathcal{M}^{\mathrm{free}} \equiv \bar{S}''[\bar{x}_\mathrm{FV}] = - \frac{d^2}{d \bar{r}^2} + 1,
\end{equation}
where $\bar{r} = \abs{\bar{t}}$ and
\begin{equation}
    \mathcal{V} \paren{\bar{r}} \equiv \left.\paren{\frac{d^2 \bar{V}}{d \bar{x}^2}}\right|_{\bar{x} = \bar{x}_b \paren{\bar{r}}} - 1
    .
\end{equation}
These two differential operators are both parity-conserving, so each operator has two superselection sectors: odd functions of $\bar{t}$ and even functions of $\bar{t}$. The zero mode of $\mathcal{M}$ is an odd function. We can thus break up the operators into
\begin{equation}
    \mathcal{M} = \mathcal{M}_{\mathrm{odd}} \mathcal{M}_{\mathrm{even}},
\end{equation}
\begin{equation}
    \mathcal{M}^{\mathrm{free}} = \mathcal{M}^{\mathrm{free}}_{\mathrm{odd}} \mathcal{M}^{\mathrm{free}}_{\mathrm{even}}
\end{equation}
and compute the functional determinant ratios for each sector.

All even modes have nonzero eigenvalues. From the Gel'fand-Yaglom theorem,
\begin{equation}
	\frac{\det \paren{\mathcal{M}_{\mathrm{even}}}}{\det \paren{\mathcal{M}_{\mathrm{even}}^\mathrm{free}}} = \frac{\psi_{\mathrm{even}}\paren{\infty}}{\psi_{\mathrm{even}}^\mathrm{free}\paren{\infty}} = \mathcal{R}_{\mathrm{even}} \paren{\infty}
	,
\label{eq:det_M_even}
\end{equation}
where $\psi_{\mathrm{even}}$ and $\psi_{\mathrm{even}}^\mathrm{free}$ are regular solutions of
\begin{equation}
	\mathcal{M}_{\mathrm{even}} \psi_{\mathrm{even}}  = 0
	,
\end{equation}
\begin{equation}
	\mathcal{M}_{\mathrm{even}}^\mathrm{free} \psi_{\mathrm{even}} ^\mathrm{free} = 0
	,
\end{equation}
and
\begin{equation}
	\mathcal{R}_{\mathrm{even}} \paren{\bar{r}} \equiv \frac{\psi_{\mathrm{even}}\paren{\bar{r}}}{\psi_{\mathrm{even}}^\mathrm{free}\paren{\bar{r}}}
	.
\label{eq:det_ratio_even}
\end{equation}
After some algebra, we obtain the equation for $\mathcal{R}_{\mathrm{even}}$,
\begin{equation}
	\mathcal{R}_{\mathrm{even}}'' \paren{\bar{r}} + 2 \tanh \paren{\bar{r}} \mathcal{R}_{\mathrm{even}}' \paren{\bar{r}} - \mathcal{V} \paren{\bar{r}} \mathcal{R}_{\mathrm{even}} \paren{\bar{r}} = 0
\end{equation}
with the initial condition $\mathcal{R}_{\mathrm{even}} (0) = 1$ and $\mathcal{R}_{\mathrm{even}}' (0) = 0$.
This is an ordinary differential equation that can be solved numerically once the exact form of the potential is given. Then we take the limit $\bar{r} \rightarrow \infty$ to compute $\mathcal{R}_{\mathrm{even}} (\infty)$.

We cannot use the same method to compute the determinant ratio in the odd sector because of the zero mode. Instead, we apply the collective coordinate method to systematically remove the zero mode~\cite{PhysRevD.72.125004}. The result is
\begin{equation}
	\paren{\frac{\overline{S_b}}{2 \pi}}^{1/2}  \paren{ \frac{\det' \paren{\mathcal{M}_{\mathrm{odd}}}}{\det \paren{\mathcal{M}_{\mathrm{odd}}^\mathrm{free}}} }^{-1/2} = \sqrt{\frac{ - \bar{x}_\infty \bar{x}_b'' \paren{0} }{\pi}}
	,
\label{eq:det_ratio_odd}
\end{equation}
where $\bar{x}_\infty$ is defined by the asymptotic behavior of $\bar{x}_b$ at $r \rightarrow \infty$
\begin{equation}
	\bar{x}_b \paren{\bar{r}} \approx \bar{x}_\infty e^{-\bar{r}}
	.
\end{equation}
Combining Eqs.~(\ref{eq:Gamma_GY_dimensionless}), (\ref{eq:det_ratio_even}), and (\ref{eq:det_ratio_odd}) we obtain
\begin{equation}
\begin{aligned}
	\Gamma = m \sqrt{\beta} e^{-\beta \overline{S_b}} \sqrt{\frac{ - \bar{x}_\infty \bar{x}_b'' \paren{0} }{\pi}} \mathrm{Im} \bracket{ \mathcal{R}_{\mathrm{even}}\paren{\infty}^{-1/2}}
	,
\end{aligned}
\label{eq:gamma_nlo_gy_qm}
\end{equation}
where $\mathcal{R}_{\mathrm{even}} (\infty)$ is negative. 

\section{Specification of \texorpdfstring{$S_V^b$}{}-cut\label{sec:sbv}}

In Sec.~\ref{sec:cut} we introduced the quantity $S^b_V$ defined on each MC configuration and used two cuts on it (ensemble-generation-level and postselection) to prevent sampling  problematic configurations that probe too close to the true vacuum. The postselection cut was placed at the minimum of the probability density of configurations as a function of $S^b_V$,
\begin{equation}
    p \paren{S^b_V} = \left| \frac{d}{d S^b_V} \paren{ \frac{\int_{S^b_V\bracket{x} > S^b_V} \mathcal{D} x \, e^{- S \bracket{x}}}{\int_{S^b_V\bracket{x} > \paren{S^b_V}^{\mathrm{min}}} \mathcal{D} x \, e^{- S\bracket{x}}}} \right|
    .
    \label{eq:pSbVappx}
\end{equation}
In this appendix we discuss the properties of $p \paren{S^b_V}$ in more detail and give a physical model to explain the typical finding ${S^b_V}|_{{\rm min}(p \paren{S^b_V})}\approx -1/2$.

The denominator in Eq.~(\ref{eq:pSbVappx}) is independent of $S^b_V$ and serves as a normalization factor for the total probability such that $\int_{(S^b_V)^{\mathrm{min}}}^{0} d S^b_V \, p (S^b_V) = 1$. We denote
\begin{equation}
    \int_{S^b_V\bracket{x} > \paren{S^b_V}^{\mathrm{min}}} \mathcal{D} x \, e^{- S\bracket{x}} = \mathcal{N}^{-1}
\end{equation}
for simplicity.
We define the density of number of configurations per $S_V^b$ as
\begin{equation}
    D \paren{S_V^b} = \lim_{\Delta S_V^b \rightarrow 0} \frac{\int_{S^b_V < S^b_V\bracket{x} < S^b_V + \Delta S_V^b} \mathcal{D} x}{\Delta S_V^b}
    ,
\end{equation}
and the average value of $e^{-S}$ over configurations conditional on $S_V^b$ as
\begin{equation}
    \expval{e^{-S}}_{S_V^b} = \lim_{\Delta S_V^b \rightarrow 0} \frac{\int_{S^b_V < S^b_V\bracket{x} < S^b_V + \Delta S_V^b} \mathcal{D} x \, e^{- S \bracket{x}}}{\int_{S^b_V < S^b_V\bracket{x} < S^b_V + \Delta S_V^b} \mathcal{D} x }
    .
\end{equation}
Then the probability density of configurations per $S_V^b$ can be rewritten as
\begin{equation}
    p \paren{S^b_V} = \mathcal{N} D \paren{S_V^b} \expval{e^{-S}}_{S_V^b}
    .
\label{eq:p_SbV_0}
\end{equation}
Qualitatively speaking, $D (S_V^b)$ is an increasing function with $D (S_V^b = 0) = \infty$ because of the enormous number of configurations with $S_V^b = 0$. $\expval{e^{-S}}_{S_V^b}$ may be a decreasing function of $S_V^b$, especially when $S_V^b$ is very low and dominates the change in the total action $S$, so that $\expval{e^{-S}}_{S_V^b} = e^{-S_V^b} \expval{e^{-(S - S_V^b)}}_{S_V^b} \approx \mathrm{constant} \times e^{-S_V^b}$. Because of the opposite monotonicities of the two factors, $p (S_V^b)$ may have a minimum.

\begin{figure}[!t]
\centering
	\begin{subfigure}[ht]{0.48\textwidth}
		\centering
\includegraphics[width=\textwidth]{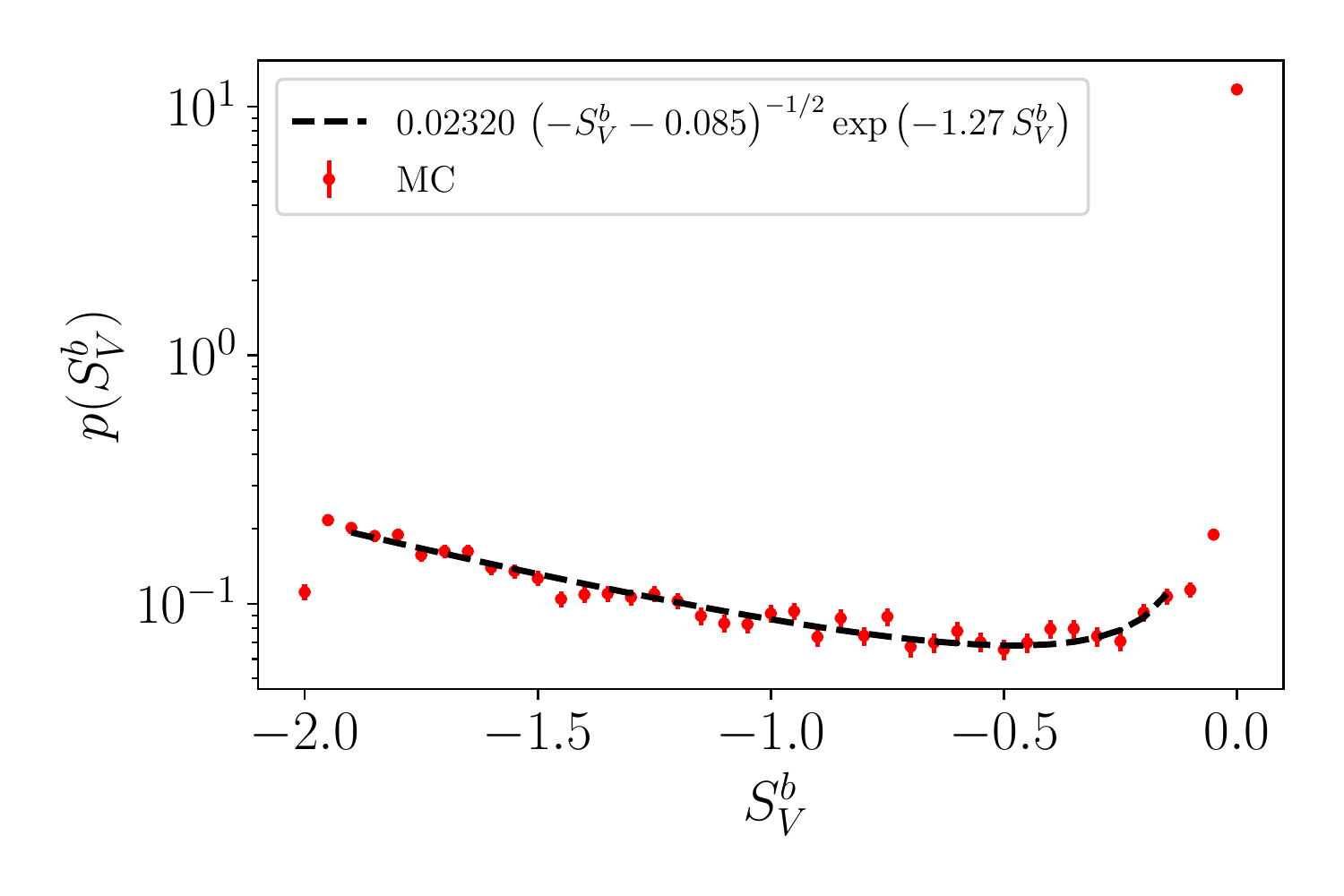}
		\caption{}
		\label{subfig:configDens_vs_SbV_1}
	\end{subfigure}
	\hfill
	\begin{subfigure}[!htbp]{0.48\textwidth}
		\centering
		\includegraphics[width=\textwidth]{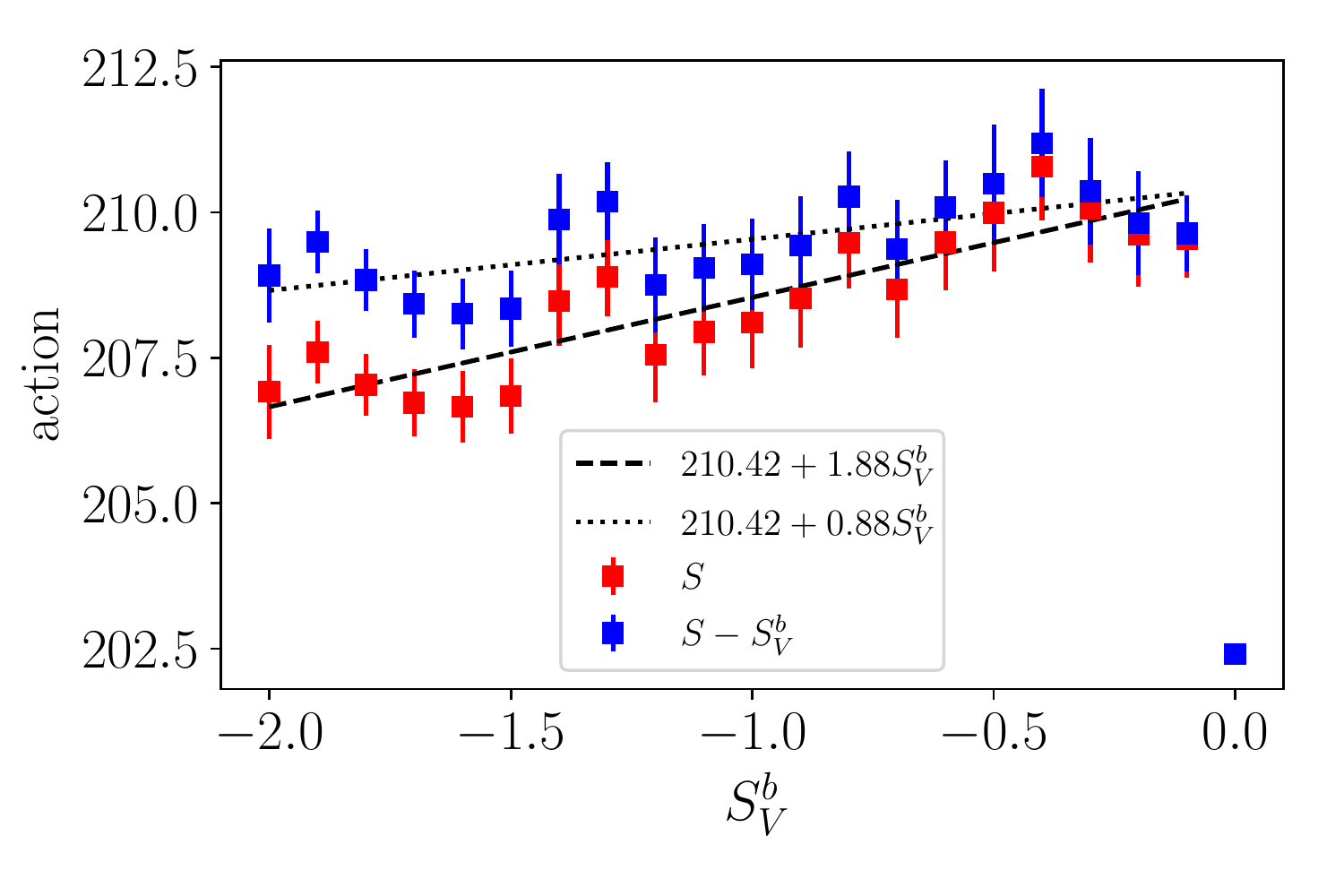}
		\caption{}
		\label{subfig:action_vs_SbV_1}
	\end{subfigure}
	\begin{subfigure}[ht]{0.48\textwidth}
		\centering
		\includegraphics[width=\textwidth]{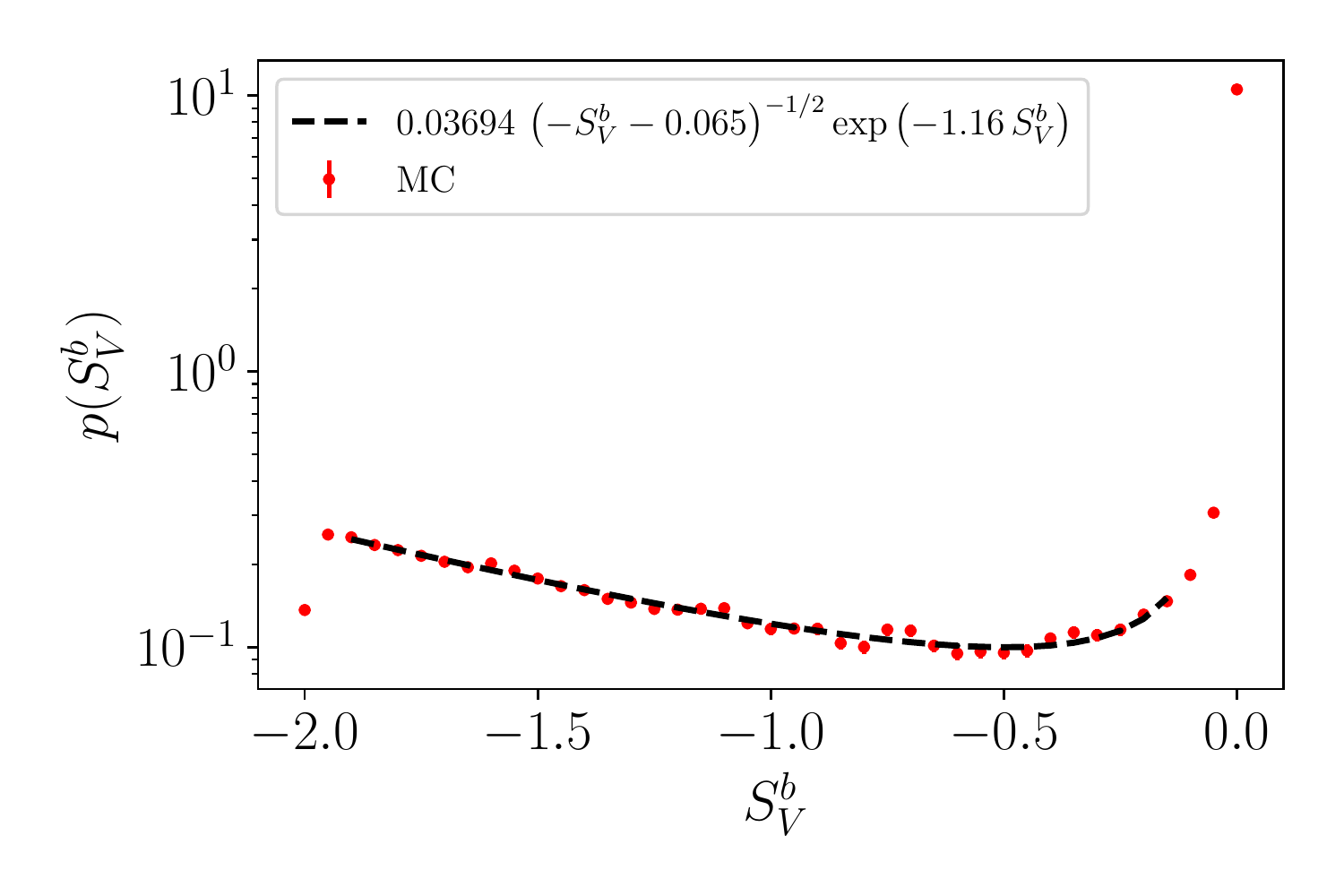}
		\caption{}
		\label{subfig:configDens_vs_SbV_2}
	\end{subfigure}
	\hfill
	\begin{subfigure}[!htbp]{0.48\textwidth}
		\centering
		\includegraphics[width=\textwidth]{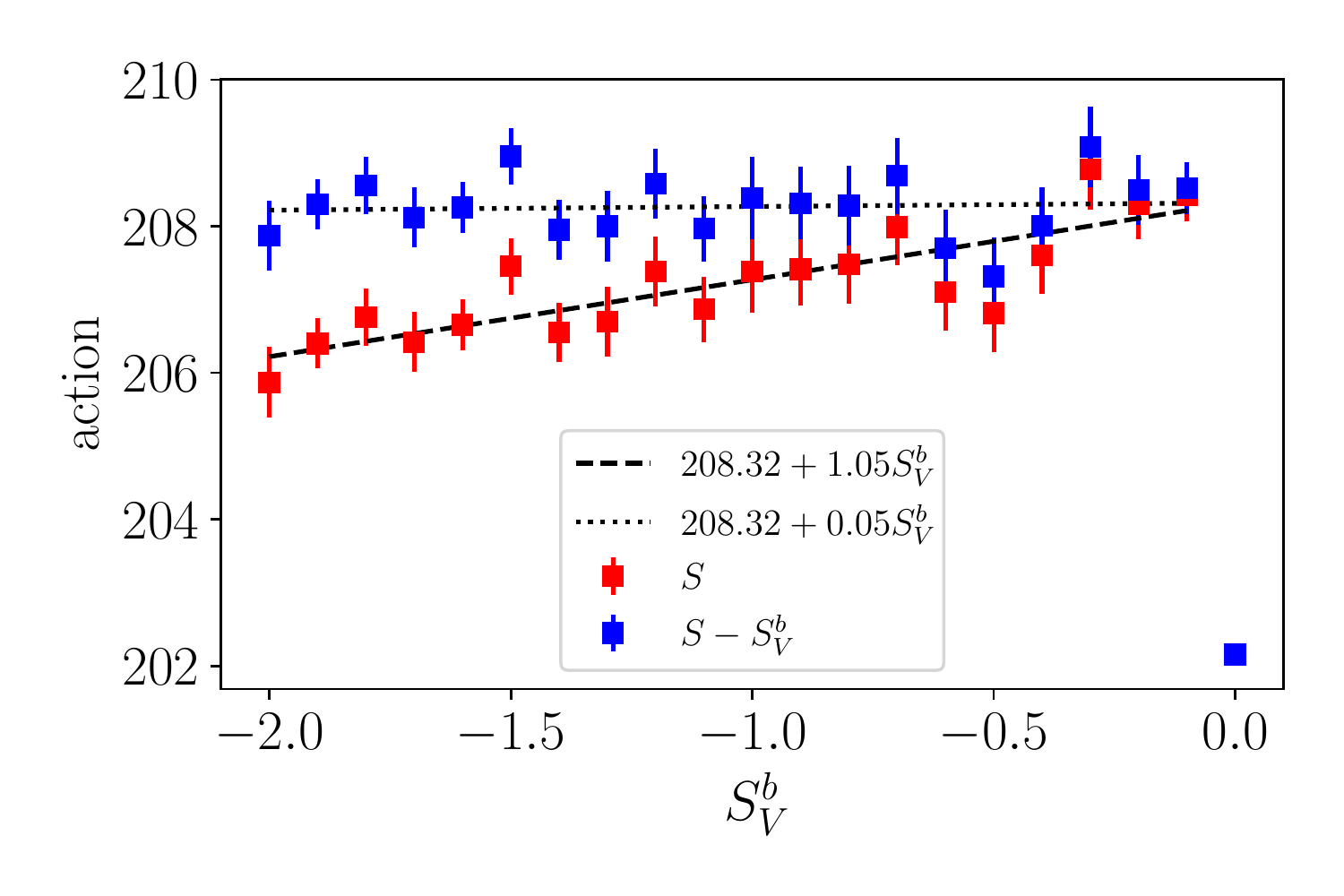}
		\caption{}
		\label{subfig:action_vs_SbV_2}
	\end{subfigure}
\caption{(\subref{subfig:configDens_vs_SbV_1})(\subref{subfig:action_vs_SbV_1}) $\alpha = 0.9$, $\beta = 8.0$, (\subref{subfig:configDens_vs_SbV_2})(\subref{subfig:action_vs_SbV_2}) $\alpha = 0.7$, $\beta = 9.0$, both with $S_V^b$ cut at $-2.0$, $a = 0.1m^{-1}$, $T = 20.0m^{-1}$. (\subref{subfig:configDens_vs_SbV_1})(\subref{subfig:configDens_vs_SbV_2}) Densities of configurations per $S_V^b$ calculated from kernel density estimation. The overall normalization of $p(S_V^b)$ is subject to the $S_V^b$ cut. Fits using Eq.~(\ref{eq:p_SbV_1}) are shown for comparison.  (\subref{subfig:action_vs_SbV_1})(\subref{subfig:action_vs_SbV_2}) Statistical dependence between $S_V^b$ and the total action $S$ calculated from kernel regression (also by using the Epanechnikov kernel with suitable choice of the kernel width). Linear fits for both $S$ and $S - S_V^b$ are shown for comparison. The linear coefficient for $S - S_V^b$ in Fig.~\ref{fig:vs_SbV}(\subref{subfig:action_vs_SbV_2}) is qualitatively close to $0$. In Fig.~\ref{fig:vs_SbV}(\subref{subfig:action_vs_SbV_1}), the linear coefficient $0.88$ is greater than $0$, but this discrepancy is comparable to the generic statistical uncertainty at each point.
}
\label{fig:vs_SbV}
\end{figure}

As is shown in Figs.~\ref{fig:vs_SbV}(\subref{subfig:action_vs_SbV_1}) and \ref{fig:vs_SbV}(\subref{subfig:action_vs_SbV_2}), for relatively large values of $\abs{S_V^b}$, the total action approximately obeys $S = S_V^b + \mathrm{constant}$. This observation supports the expectation described in the previous paragraph that $\expval{e^{-S}}_{S_V^b} \approx \mathrm{constant} \times e^{-S_V^b}$. We fit the curves over a range of $S^b_V$ chosen by hand to demonstrate the idea. The fit is not used for computation of the final result of the decay rate. Statistical errors from Monte Carlo are not considered in the fit for simplicity. 

Further, with decreasing $S_V^b$, the total action $S$ is approximately decreasing. The negative mode with $\lambda_0 < 0$ is the only mode that lowers the total action when going away from the bounce solution. Therefore, in this region, the change in $c_0$ dominates the change in the total action and also the change in $S_V^b$.
Under this assumption, we have $S_V^b \approx (S_V^b)_0 + (1/2) \lambda_0 (c_0)^2$ and $S \approx \mathrm{constant} + S_V^b$. $(S_V^b)_0$ is a point in the $S_V^b$-space from which $S - S_V^b$ starts to decrease with increasing $S_V^b$, \textit{i.e.}, no longer independent of the value of $S_V^b$. Then,
\begin{align}
    D \paren{S_V^b} = & \lim_{\Delta S_V^b \rightarrow 0} \frac{\int_{S^b_V < S^b_V\bracket{x} < S^b_V + \Delta S_V^b} \mathcal{D} x}{\Delta S_V^b} \nonumber\\
    = & \lim_{\Delta c_0 \rightarrow 0} \frac{\int_{c_0 < c_0 \bracket{x} < c_0 + \Delta c_0} \mathcal{D} x}{\left|\lambda_0 c_0 \right| \Delta c_0} \nonumber\\
    = & \mathrm{constant} \times \frac{1}{\left|\lambda_0 c_0 \right|} \nonumber\\
    = & \mathrm{constant} \times \frac{1}{\sqrt{2 \left|\lambda_0 \paren{S_V^b - (S_V^b)_0} \right|}}
\label{eq:D_SbV_0}
\end{align}
where $c_0 [x]$ is the negative mode coefficient of the configuration $x(t)$, and $\int_{c_0 < c_0 [x] < c_0 + \Delta c_0} \mathcal{D} x = \mathrm{constant} \times \Delta c_0$. Combining these observations we obtain an approximate model for the probability density, 
\begin{equation}
    p \paren{S^b_V} = \mathrm{constant} \times \frac{1}{\sqrt{- \paren{S_V^b - (S_V^b)_0}}} e^{-S_V^b}
    .
\label{eq:p_SbV_1}
\end{equation}
The minimum is
\begin{equation}
    0=\frac{1}{p \paren{S^b_V}} \frac{d p \paren{S^b_V}}{d S_V^b} = \frac{1}{D \paren{S_V^b}} \frac{d D \paren{S_V^b}}{d S_V^b} + \frac{1}{\expval{e^{-S}}_{S_V^b}} \frac{d \expval{e^{-S}}_{S_V^b}}{d S_V^b} = -\frac{1}{2 \paren{S_V^b - (S_V^b)_0}} - 1
\end{equation}
or
\begin{align}
    (S_V^b)_{min} = (S_V^b)_0 - 1/2
    .
\end{align}
Physically we expect $(S_V^b)_0$ to be small, and approximating $(S_V^b)_0\approx 0$ gives the minimum $(S_V^b)_{min}=-1/2$.

Now let us compare with Monte Carlo.
In  Figs.~\ref{fig:vs_SbV}(\subref{subfig:configDens_vs_SbV_1}) and \ref{fig:vs_SbV}(\subref{subfig:configDens_vs_SbV_2}), we examine results from two simulated potentials and we fit the measured $p(S^b_V)$ with a model similar to (but slightly generalizing) Eq.~(\ref{eq:p_SbV_1}). The fit is not used for the computations of the decay rate, only for the illustration of the physics of the quantity $S^b_V$.  There is some subjectivity in choosing the fit range of $S^b_V$,  because the lower end of the MC result is affected by the cut on $S^b_V$, and the upper end of $S^b_V \approx 0$ is not expected to satisfy the conditions for the above arguments. Statistical errors in the density of configurations from Monte Carlo are not considered in the fit for simplicity.

Our argument for the functional form of $p(S^b_V)$ is not meant to be precise. We see that the model fit is good, but there are deviations from Eq.~(\ref{eq:p_SbV_1}). For example, the  coefficient in the exponent returned by the fits is not exactly $-1$. The constant $(S_V^b)_0$, in the example of Fig.\ref{fig:vs_SbV}~(\subref{subfig:action_vs_SbV_1}), is about $-0.1$. The fit in  Fig.~\ref{fig:vs_SbV}(\subref{subfig:configDens_vs_SbV_1}) gives $(S_V^b)_0 \approx -0.085$.  Similar inaccuracies in the model can also be seen in Figs.~\ref{fig:vs_SbV}(\subref{subfig:action_vs_SbV_2}) and \ref{fig:vs_SbV}(\subref{subfig:configDens_vs_SbV_2}). However, it suffices as a qualitative description, and indeed we find in our numerical studies that the stationary point of $p(S_V^b)$ is generically in the range $-1.0$ to $-0.1$. The most important conclusion is that it is reasonable to expect the probability density to have {\emph{a}} minimum, roughly somewhere in this range.

We now use semiclassical arguments to assert that the effect of varying the cut on $S_V^b$, near the stationary point of $p(S^b_V)$, results only in an $O(1)$ uncertainty in the decay rate.  If we define the cut as $(S_V^b)^{\mathrm{min}} = (S_V^b)_0-1/2$, then $(1/2) |\lambda_0|  (c_0^{\mathrm{max}})^2 = 1/2$. Combined with the previously discussed cut $(1/2) |\lambda_0|  (c_0^{\mathrm{min}})^2 = 1$, the $c_0$ integral in the semiclassical computation is
\begin{align}
\int_{c_0^{\mathrm{min}}}^{c_0^{\mathrm{max}}} dc_0\, e^{-\frac{1}{2} \lambda_0 c_0^2} &= \sqrt{\frac{\pi}{2 \left|\lambda_0\right|}} \bracket{\mathrm{Erfi}\paren{\sqrt{\frac{\left| \lambda_0 \right|}{2}} c_0^{\mathrm{max}}  } + \mathrm{Erfi}\paren{\sqrt{\frac{\left| \lambda_0 \right|}{2}} \left|c_0^{\mathrm{min}}\right|}}\nonumber\\
&= \sqrt{\frac{\pi}{2 \left|\lambda_0\right|}} \paren{\mathrm{Erfi}\paren{\frac{1}{\sqrt{2}}} + \mathrm{Erfi}\paren{1}} \nonumber\\
&= 1.30193 \sqrt{2 \frac{\pi}{ \left|\lambda_0\right|}} \nonumber\\
&= 1.30193 \, \mathrm{Im} \int_{- i \infty}^{i \infty} dc_0\, e^{-\frac{1}{2} \lambda_0 c_0^2}
,
\end{align}
which differs from the result from analytic continuation only by an $O(1)$ factor.

To summarize, we propose to place a cut the configurations at the value of $S^b_V$ at the minimum of the sample distribution $p(S^b_V)$. In practical Monte Carlo simulations, a lower cut in $S^b_V$ that contains the stationary point is needed in ensemble generations in order to find the appropriate cut in $S^b_V$. A relatively small ensemble may be enough for giving a conservative estimation of where to cut. Then, a postselection of configurations discards configurations with $S^b_V$ lower than the stationary point. Computation of observables is then performed on the ensemble after postselection.

\section{Details of the numerical computations \label{sec:details_computation}}

In this appendix we provide  details of the MC ensembles and the methods used to numerically analyze the MC data.

\subsection{Parameters of the ensembles}

\begin{table}[!ht]
\begin{center}
\footnotesize
\centering
\begin{tabular}{c c c c c c c c c}
\hline\hline
Figure & \begin{tabular}[x]{@{}c@{}}$a$\\(units of $m^{-1}$)\end{tabular} & $N_T$ & $S_V^b$-cut & $\alpha$ & $\beta$ & $N_{\mathrm{cf}}$ & \begin{tabular}[x]{@{}c@{}}Postselection \\ $S_V^b$-cut\end{tabular} & $N_{\mathrm{cf,post}}$ \\
\hline
 Figure~\ref{fig:vary_beta} & $0.1$ & $400$ & $-2.0$ &$0.9$ & $4.0$ & $$19{,}999$$ & $-0.2879$ & $1{,}864$ \\
 varying $\beta$ & $0.1$ & $400$ & $-2.0$ &$0.9$ & $50$ & $20{,}000$ & $-0.3304$ & $3{,}594$ \\
 & $0.1$ & $400$ & $-1.0$ &$0.9$ & $6.0$ & $20{,}000$ & $-0.5010$ & $15{,}259$ \\
 & $0.1$ & $400$ & $-2.0$ &$0.9$ & $7.0$ & $19{,}999$ & $-0.5672$ & $11{,}390$ \\
 & $0.1$ & $400$ & $-2.0$ & $0.9$ & $8.0$ & $20{,}000$ & $-0.4966$ & $16{,}487$ \\
 & $0.1$ & $400$ & $-2.0$ & $0.9$ & $9.0$ & $40{,}000$ & $-0.6261$ & $36{,}027$ \\
 & $0.1$ & $400$ & $-2.5$ & $0.9$ & $10.0$ & $59{,}999$ & $-0.5293$ & $55{,}697$ \\
 & $0.1$ & $400$ & $-2.5$ & $0.9$ & $11.0$ & $119{,}999$ & $-0.6925$ & $117{,}888$ \\
 \hline
 Figure~\ref{fig:vary_alpha} & $0.1$ & $400$ & $-2.0$ & $0.6$ & $9.0$ & $40{,}000$ & $-0.5385$ & $27{,}819$ \\
 varying $\alpha$ & $0.1$ & $400$ & $-2.0$ & $0.7$ & $9.0$ & $40{,}000$ & $-0.5406$ & $30{,}747$ \\
 & $0.1$ & $400$ & $-2.0$ & $0.8$ & $9.0$ & $40{,}000$ & $-0.5473$ & $32{,}544$ \\
 & $0.1$ & $400$ & $-2.0$ & $0.9$ & $9.0$ & $40{,}000$ & $-0.6261$ & $36{,}027$ \\
 & $0.1$ & $400$ & $-2.0$ & $0.95$ & $9.0$ & $40{,}000$ & $-0.5473$ & $37{,}811$ \\ 
 \hline
Figure~\ref{fig:vary_SbV_cut_T_a}(\subref{subfig:Gamma_vs_SbV-cut_alpha_0.900_beta_8.000}) & $0.1$ & $400$ & $-3.0$ & $0.9$ & $8.0$ & $9{,}999$ & $-0.2408$ & $3{,}688$ \\
varying $S_V^b$-cut & $0.1$ & $400$ & $-2.5$ & $0.9$ & $8.0$ & $9{,}999$ & $-0.6711$ & $6{,}775$ \\
& $0.1$ & $400$ & $-2.0$ & $0.9$ & $8.0$ & $20,000$ & $-0.4966$ & $16{,}487$ \\
& $0.1$ & $400$ & $-1.5$ & $0.9$ & $8.0$ & $9{,}999$ & $-0.4002$ & $8{,}741$ \\
\hline
Figure~\ref{fig:vary_SbV_cut_T_a}(\subref{subfig:Gamma_vs_t_alpha_0.900_beta_8.000}) & $0.1$ & $200$ & $-2.0$ & $0.9$ & $8.0$ & $20,000$ & $-0.4669$ & $15{,}620$ \\
varying $T$ & $0.1$ & $400$ & $-2.0$ & $0.9$ & $8.0$ & $20,000$ & $-0.4966$ & $16{,}487$ \\
 & $0.1$ & $600$ & $-2.0$ & $0.9$ & $8.0$ & $20,000$ & $-0.4527$ & $11{,}840$ \\
 & $0.1$ & $800$ & $-2.0$ & $0.9$ & $8.0$ & $19{,}999$ & $-0.5906$ & $12{,}685$ \\
 & $0.1$ & $1000$ & $-2.0$ & $0.9$ & $8.0$ & $20{,}000$ & $-0.4326$ & $9{,}421$ \\
\hline
Figure~\ref{fig:vary_SbV_cut_T_a}(\subref{subfig:Gamma_vs_a_t_20.00_alpha_0.900_beta_8.000}) & $0.05$ & $800$ & $-2.0$ &$0.9$ & $8.0$ & $20{,}000$ & $-0.4170$ & $15{,}043$ \\
varying $a$ & $0.1$ & $400$ & $-2.0$ & $0.9$ & $8.0$ & $20{,}000$ & $-0.4966$ & $16{,}487$ \\
 & $0.2$ & $200$ & $-2.0$ & $0.9$ & $8.0$ & $20{,}000$ & $-0.4993$ & $15{,}842$ \\
 & $0.4$ & $100$ & $-2.0$ & $0.9$ & $8.0$ & $19,999$ & $-0.4313$ & $14{,}590$ \\
\hline\hline
\end{tabular}
\end{center}
\caption{
Parameters used in ensembles. Results for the MC estimate of $\Gamma$ are given in the corresponding figures where they are compared with three alternative computation methods: the leading-order semiclassical approximation, where the dimensionful prefactor is estimated with dimensional analysis (DA), $\Gamma_{DA} = m e^{-S_b}$; the NLO semiclassical approximation using the Gel'fand-Yaglom (GY) method; and the solution of the time-dependent Schr\"odinger equation (TDSE) by exact diagonalization. $N_{\mathrm{cf}}$ is the number of configurations. (Where ensembles from different figures have same parameters,  the same ensemble is used.)
}
\label{table:ensembles}
\end{table}

Table~\ref{table:ensembles} shows the parameters of the ensembles used in Figs.~\ref{fig:vary_beta},~\ref{fig:vary_alpha},~\ref{fig:vary_SbV_cut_T_a}.

\subsection{Binning the postselected ensembles}

After analyzing the distribution of configurations in $S_V^b$-space we apply the postselection $S_V^b$-cut. The retained configurations define the postselection ensemble on which we measure observables. Since our original ensembles have autocorrelation, the postselection ensemble is also autocorrelated.
With a bin size $K$, the effective number of independent configurations is $N_{\mathrm{cf,post}} / K$.

Considering small and large limits of the ratio $N_{\mathrm{cf,post}} / N_{\mathrm{cf}}$ can be used to justify binning the Monte Carlo configurations in the postselected ensemble. In the small limit, the ratio $N_{\mathrm{cf,post}} / N_{\mathrm{cf}} \rightarrow 0$, and the postselected configurations become essentially uncorrelated. In this case, binning is unnecessary.
In the large limit, the ratio $N_{\mathrm{cf,post}} / N_{\mathrm{cf}} \rightarrow 1$, and the postselected ensemble is similar to the original ensemble, where binning is standard.

On our postselected ensembles, to determine the suitable bin sizes $K$, we change $K$ and calculate the statistical error of $\hat{\rho}$ on binned configurations (by using the mean value of $\hat{\rho}$ over each bin). For ensembles in Table~\ref{table:ensembles}, a generic suitable choice of $K$ turns out to be about $200$, where the statistical error starts to saturate. 
This bin size is used to obtain the statistical errors shown in
Figs.~\ref{fig:vary_beta},~\ref{fig:vary_alpha},~\ref{fig:vary_SbV_cut_T_a}.

\subsection{Numerical details of the constrained ensemble reweighting
\label{sec:num_details_longlifetime}}
\begin{figure}[!ht]
\centering
	\begin{subfigure}[ht]{0.48\textwidth}
		\centering
\includegraphics[width=\textwidth]{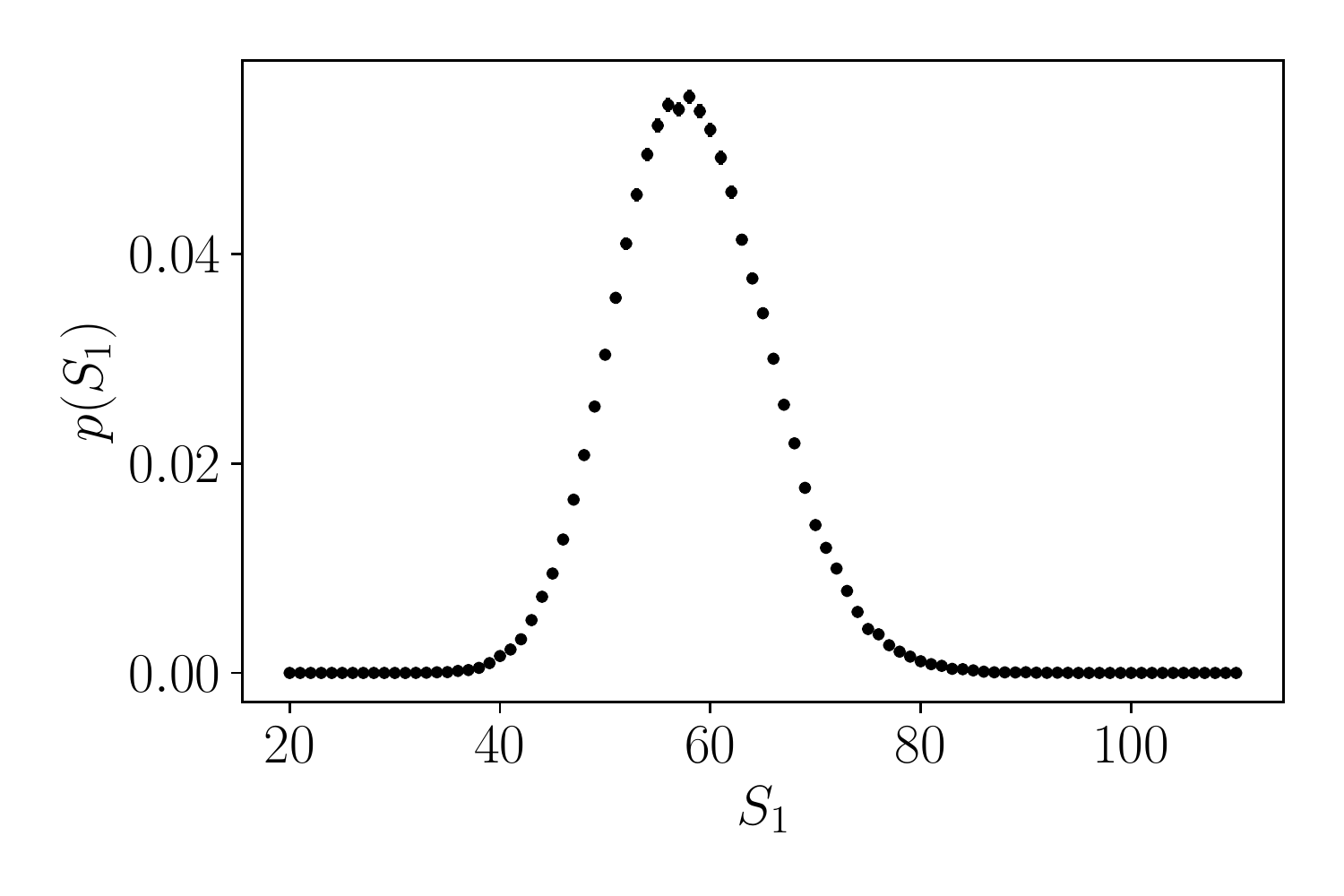}
		\caption{Distribution of configurations in ensemble 1 that satisfy $x_{\mathrm{FV}}(t_i) - \Delta x / 2 < x(t_i) < x_{\mathrm{FV}}(t_i) + \Delta x / 2$.}
		\label{subfig:long_lifetime_configDens_vs_S1}
	\end{subfigure}
	\hfill
	\begin{subfigure}[!htbp]{0.48\textwidth}
		\centering
		\includegraphics[width=\textwidth]{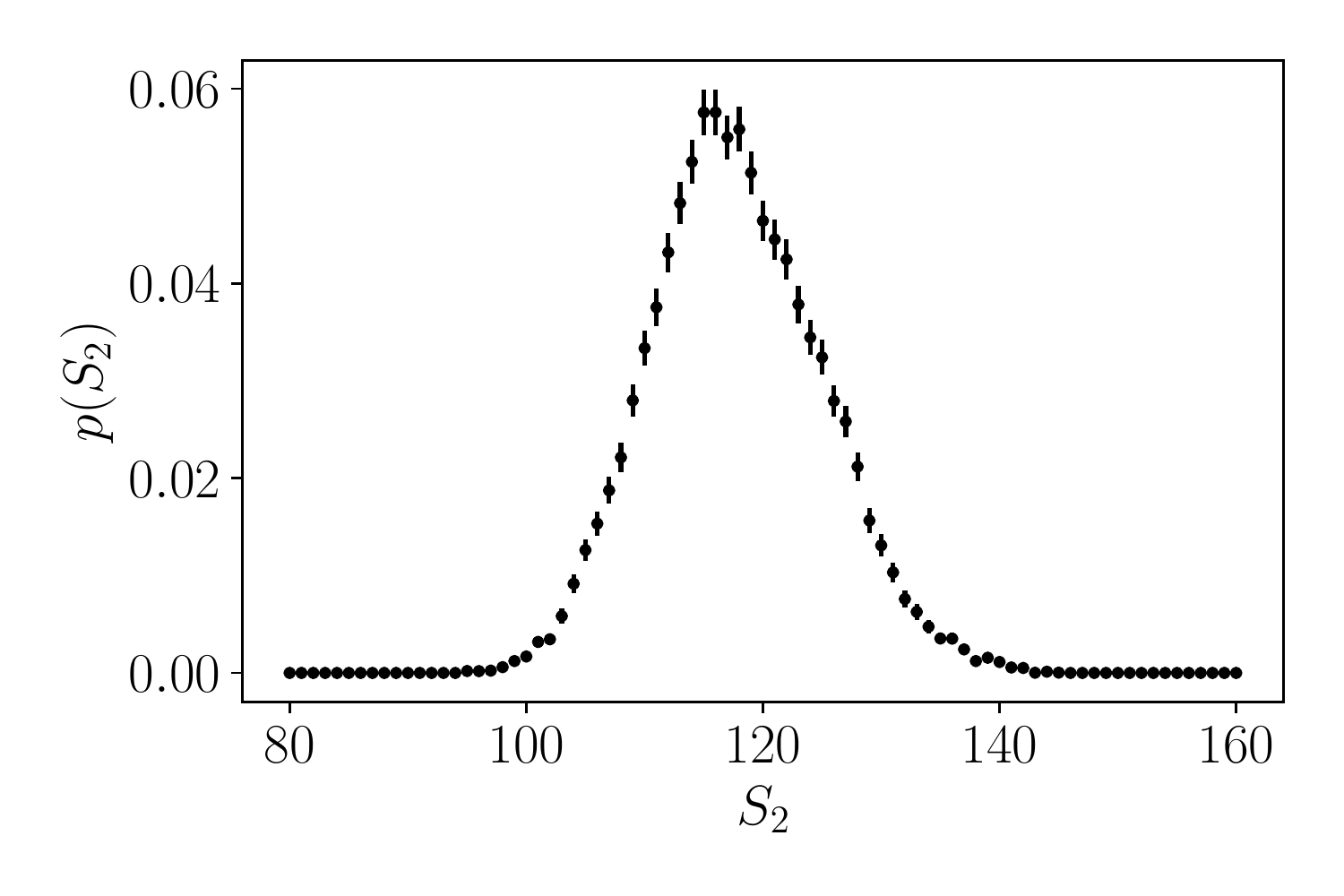}
		\caption{Distribution of configurations in ensemble 2 that satisfy $x^{\diamond}(t_i) - \Delta x / 2 < x(t_i) < x^{\diamond}(t_i) + \Delta x / 2$.}
		\label{subfig:long_lifetime_configDens_vs_S2}
	\end{subfigure}
\caption{Distributions of original configurations with $K = 5$, $\Delta x = 4.0 m^{-1/2}$ and the kernel width $h = 1.0$ in KDE using the Epanechnikov kernel. Fig.~(\subref{subfig:long_lifetime_configDens_vs_S1}): ensemble 1; Fig.~(\subref{subfig:long_lifetime_configDens_vs_S2}): ensemble 2.
}
\label{fig:long_lifetime_distribution_vs_S}
\end{figure}

When applying the constrained ensemble reweighting technique  introduced in  Sec.~\ref{sec:longlifetime}, we find that the relevant quantities used in the calculation, such as the frequency of the event $x^{\diamond}(t_i) - \Delta x / 2 < x(t_i) < x^{\diamond}(t_i) + \Delta x / 2$, are not sensitive to the bin size $K$. In practice, we use $K = 5$ as a safer choice than $K = 1$. To prevent the smoothing artifacts due to taking the average of $K$ configurations, for every $K$ configurations, we only use one configuration in the calculation and skip the remaining $K - 1$ configurations.

We show the probability distribution of original configurations in ensembles 1 and 2 in Fig.~\ref{fig:long_lifetime_distribution_vs_S} with $\Delta x = 4.0 m^{-1/2}$.
We find that $p(S_1)$ and $p(S_2)$ are peaked at much higher values of $S_1$ and $S_2$ than the actions of the median smoothed configurations (about $0.142$ for ensemble 1 and about
$62.082$ for ensemble 2). As explained above this is to be expected due to high-frequency fluctuations in the original configurations. 
In calculation of Sec.~\ref{sec:longlifetime}, we use the statistics of $S_1$ and $S_2$ over the distributions (also subject to the change in $\Delta x$) at variable $\Delta x$ and use them jointly to evaluate $e^{- S\bracket{x^{\diamond}} + S\bracket{x_{\mathrm{FV}}}}$ in Eq.~(\ref{eq:c1_over_c2}).

\end{appendices}

\renewcommand\refname{References}
\bibliographystyle{unsrt}
\bibliography{main}

\end{document}